\documentclass{emulateapj}
\usepackage{amsmath}
\usepackage{graphicx}
\usepackage{amsfonts}
\usepackage{natbib}
\usepackage{color}

\newcommand{\al}{\alpha}
\newcommand{\be}{\beta}
\newcommand{\gam}{\gamma}
\newcommand{\Gam}{\Gamma}

\newcommand{\De}{\Delta}
\newcommand{\eps}{\epsilon}
\newcommand{\sig}{\sigma}
\newcommand{\Sig}{\Sigma}
\newcommand{\ka}{\kappa}

\newcommand{\Om}{\Omega}
\newcommand{\Mach}{\mathcal{M}}
\newcommand{\pd}{\partial}
\newcommand{\dd}{\mbox{d}}

\newcommand{\Disco}{{\texttt{Disco}}}
\newcommand{\model}[1]{{Model \texttt{#1}}}

\newcommand{\OO}{\mathcal{O}}

\newcommand{\ave}[1]{\left \langle #1 \right \rangle}
\newcommand{\avet}[1]{ \langle #1 \rangle}
\newcommand{\aveRe}[1]{\left \langle #1 \right \rangle_\text{Re}}

\begin{document}

\title{Minidisks in Binary Black Hole Accretion}
\author{Geoffrey Ryan\altaffilmark{1,a} and Andrew MacFadyen\altaffilmark{1}}
\altaffiltext{a}{gsr257@nyu.edu}
\altaffiltext{1}{Center for Cosmology and Particle Physics, Physics Department, New York University, New York, NY 10003, USA}

\begin{abstract}

Newtonian simulations have demonstrated that accretion onto binary black holes produces accretion disks around each black hole (``minidisks''), fed by gas streams flowing through the circumbinary cavity from the surrounding circumbinary disk. We study the dynamics and radiation of an individual black hole minidisk using 2D hydrodynamical simulations performed with a new general relativistic version of the moving-mesh code Disco. We introduce a comoving energy variable that enables highly accurate integration of these high Mach number flows. Tidally induced spiral shock waves are excited in the disk and propagate through the innermost stable circular orbit, providing a Reynolds stress that causes efficient accretion by purely hydrodynamic means and producing a radiative signature brighter in hard X-rays than the Novikov--Thorne model. Disk cooling is provided by a local blackbody prescription that allows the disk to evolve self-consistently to a temperature profile where hydrodynamic heating is balanced by radiative cooling. We find that the spiral shock structure is in agreement with the relativistic dispersion relation for tightly wound linear waves. We measure the shock-induced dissipation and find outward angular momentum transport corresponding to an effective alpha parameter of order 0.01. We perform ray-tracing image calculations from the simulations to produce theoretical minidisk spectra and viewing-angle-dependent images for comparison with observations.

\end{abstract}


\section{Introduction}
\label{sec:intro}

Supermassive black hole binaries (SMBHBs) are expected to form after mergers of galaxies during hierarchical structure formation.  The binaries settle to the center of the merger remnant and are expected to be embedded in gas. Gaseous accretion onto black hole binaries is thus a problem of compelling interest for understanding the dynamics and radiation of merging black holes. Pulsar timing array (PTA) measurements have started to come in tension with expected rates of nanohertz gravitational wave (GW) emission from predicted populations of SMBHBs \citep{Shannon15}. This raises the question of whether SMBHBs merge at all or are driven through the PTA band by interaction with ambient gas. Electromagnetic (EM) or GW detection of an SMBHB could help answer this question but requires a detailed understanding of the complex dynamics of these systems. \cite{Graham15A} have recently claimed that a 5.2 yr periodicity detected in Catalina Real-Time Transient Survey observations of the quasar PG 1302 corresponds to a black hole binary with mass $10^{8-9}M_{\odot}$ and separation $\sim 10^{-2}$ pc \citep[see also][]{Graham15B}. 

Black hole binaries also form as the evolutionary endpoint of massive star binaries or by capture in dense stellar environments. The recent LIGO detection of GWs from stellar-mass binary black holes calls attention to the dynamics of binary evolution in this mass range. In particular, if gas is present during some phase of these systems, circumbinary accretion will occur and may effect the orbital evolution or lead to the production of an electromagnetic counterpart \citep{Bartos16, Perna16}

Analytic treatments \citep{Artymowicz94,Milos05,Shapiro10} predicted that binary black holes with sufficiently large mass ratio embedded in gas disks would reside in a circumbinary cavity of radius twice the binary separation $a$ maintained by tidal torques. It was thought that these torques prevented accretion onto the black holes. However, multidimensional numerical simulations \citep{MacFadyen08, Noble12, Farris12, DOrazio12,Gold14, Farris14, Farris15A, Farris15B,Shi15, Bankert15,Schnittman15,delValle15,Young15,DOrazio16, Munoz16,Miranda16} have demonstrated that gas streams enter the circumbinary cavity and feed accretion disks around each of the individual black holes. We term these disks ``minidisks.'' These simulations find that accretion is not significantly suppressed compared to the accretion rate expected for a single black hole with the binary mass \citep{Farris14,Shi15}.

Electromagnetic emission from the minidisk may be of importance for identifying SMBHBs through the spectral energy distribution (SED) of observed active galactic nuclei (AGNs). \cite{Roedig14} predict a notch in the SED appearing between characteristic photon energy corresponding to the circumbinary disk and the minidisks. It is therefore of importance to carefully calculate minidisk emission models as searches for SMBHBs continue \citep{Runnoe15,Li16,Charisi16}.

Due to the large dynamic range of length scales required for this problem, global simulations of these disks have been restricted to those that either excise the cavity completely or employ mass sinks with approximate accretion prescriptions.  This is necessary to prevent artificial accumulation of mass near each black hole, but it wipes out all detailed structure of the minidisks themselves.

In this work we present the results of 2D inviscid general relativistic hydrodynamic (GRHD) simulations of accretion disks around an individual member of a black hole binary. These simulations focus on the minidisks seen in global circumbinary accretion simulations and can be seen as ``zoomed-in'' simulations of the hydrodynamics in the immediate vicinity of one of the black holes.  These simulations serve a double purpose. First, they provide a much better resolved view of minidisk emission structure, helping to inform searches for EM counterparts of SMBH binaries.  Second, the detailed accretion dynamics can inform the accretion prescriptions used in large-scale Newtonian simulations, facilitating the approach to a global understanding of circumbinary accretion.

Our GRHD simulations utilize Kerr--Schild coordinates, which have the advantage that the inner edge of the grid can be extended inside the event horizon of the black hole, thus providing a physically realistic inner boundary condition. In this study we consider the case of nonrotating black holes in the Schwarzschild metric.

  We find that minidisks accrete via ideal hydrodynamical processes alone without the need for outward angular momentum transport due to the magnetorotational instability (MRI) often modeled with an $\alpha$ prescription.
This is due to the presence of spiral shocks excited by the tidal forces of the binary companion. These shocks heat the disk and provide an outward angular momentum flux. We include local 
blackbody cooling with electron-scattering opacity to remove shock-generated
heat self-consistently from the disk.  This allows the disk to find a natural
temperature equilibrium and allows a direct estimate of the SED.

The role of spiral shocks in transporting angular momentum has been a matter of discussion for many decades.  Early numerical work by \cite{Sawada86} and analytical work by \cite{Spruit87} established the general picture: tidal forces from a binary companion excite spiral density waves that carry negative angular momentum and can steepen into shocks.  The torque from spiral shocks decreases with the scale height of the disk (or equivalently with increasing disk Mach number $\mathcal{M}$) and is sensitive to the adiabatic index $\Gamma$ of the gas.  Later numerical work confirmed these trends \citep{Godon98, Blondin00}.  Recently, \cite{Rafikov16} has established the torque-shock dissipation connection under a more general framework.  \cite{Ju16} have examined accretion due to spiral shocks in cataclysmic variables (CVs) with 2D Newtonian hydrodynamics and 3D Newtonian magnetohydrodynamics, \cite{Zhu16} performed a similar analysis with blackbody cooling in circumplanetary disks, and \cite{Bae16} have investigated the stability of spiral shocks in 3D Newtonian hydrodynamics.

In environments with sufficiently hot gas or extreme mass ratio binaries, it is possible that a cavity does not form within the circumbinary disk \citep{delValle14,delValle15}. In such conditions the accretion near the black holes will have very different structure than the stream-fed minidisk picture under consideration here. This work focuses on the case where the circumbinary disk is sufficiently thin that a cavity has formed.

This paper is organized as follows: In Section \ref{sec:numerics} we present the
numerical setup used in the simulations---a version of the \Disco{} code 
modified to work in an arbitrary spacetime, with optimizations for thin 
relativistic accretion disks including a carefully chosen energy variable.  In Section \ref{sec:models} we detail the 
minidisk models calculated.  Section \ref{sec:analysis} introduces the fiducial run and details the analysis performed.  Section \ref{sec:results} applies the analysis to all models, shows the effect of shocks on angular momentum transport, and calculates effective $\alpha$ values and spectra. Results are discussed in Section \ref{sec:discussion} and the work is summarized in Section \ref{sec:summary}.


\section{Numerical Setup}
\label{sec:numerics}

The basis of our hydrodynamics scheme is the \Disco{} code, a moving-mesh hydrodynamics
code optimized for disk geometry. This code was first used in the context of
protoplanetary disks \citep{Duffell12, Duffell13, Duffell14} and later applied 
to circumbinary accretion \citep{Farris14, Farris15A, Farris15B}. 

In the present work we have extended the \Disco{} code to solve the GRHD equations in a fixed spacetime:
\begin{equation}
    \nabla_\mu \rho_0 u^\mu = 0 \text{ and } \nabla_\mu T^{\mu\nu} = -\dot{Q} u^\nu , \label{eq:GRHD}
\end{equation}
for a single species gas of rest-mass density $\rho_0$, four-velocity $u^\mu$, stress energy tensor $T^{\mu\nu}$ and local isotropic cooling $\dot{Q}$.  

To solve Equation \eqref{eq:GRHD} numerically, one must make a choice of which elements of $T^{\mu\nu}$ to be independent variables.  We follow the standard Valencia formulation (\citealt{Marti91, Banyuls97,Font08} and implemented in, e.g. \citealt{HARM} and \citealt{Duez05}) for the momentum variables $T^0_i$ and choose an energy variable projected onto an analytically specified four-velocity $U^\mu$: $-U_\mu T^{\mu 0}$.
In terms of coordinate derivatives Equation \eqref{eq:GRHD} takes the standard flux-balanced conservation form
\begin{equation}
    \pd_0 \mathcal{U} + \pd_j \mathcal{F}^j = \mathcal{S} , \label{eq:consLaw}
\end{equation}
with conserved variables
\begin{equation}
    \mathcal{U} = \begin{pmatrix} D \\
                            S_i \\
                            \tau_U
                \end{pmatrix} = \sqrt{-g} \begin{pmatrix} \rho_0 u^0 \\ 
                                                    T^0_i \\
                                                    -U^\mu T_\mu^0 - \rho_0 u^0 \end{pmatrix} , \label{eq:cons}
\end{equation}
fluxes
\begin{equation}
    \mathcal{F}^j = \sqrt{-g} \begin{pmatrix} \rho_0 u^j \\
                                                T^j_i \\
                                                -U^\mu T_\mu^j \end{pmatrix} ,\label{eq:fluxes}
\end{equation}
and source terms 
\begin{equation}
    \mathcal{S} = \sqrt{-g} \begin{pmatrix} 0 \\
                        \frac{1}{2}T^{\mu\nu}\pd_i g_{\mu\nu} - \dot{Q}u_i \\
                        T^{\mu\nu}\nabla_\mu U_\nu + U^\mu u_\mu \dot{Q} \end{pmatrix} .\label{eq:sources}
\end{equation}

In this work we assume an ideal gas with stress tensor
\begin{equation}
	T^{\mu\nu} = \rho_0 h u^\mu u^\nu + P g^{\mu\nu} ,
\end{equation}
where $P$ is the gas pressure, $h = 1 + \eps + P/\rho_0$ is the relativistic specific enthalpy, and $\eps$ is the specific internal energy. Furthermore, we assume the gamma-law equation of state
\begin{equation}
	P = (\Gam - 1) \rho_0 \eps , \label{eq:gammalaw}
\end{equation}
where the adiabatic index $\Gamma$ is chosen to be 5/3 in accordance with \cite{Farris14}.

\Disco{} is a Godunov-type code that solves hyperbolic systems of equations of the form \eqref{eq:consLaw} on a moving mesh in cylindrical coordinates ($r$, $\phi$, $z$).  The mesh motion is restricted to be in the $\phi$-direction, which greatly reduces numerical viscosity and advection errors due to bulk azimuthal flow.  In this work, the velocities of cell interfaces are fixed to $V^\phi \equiv U^\phi/U^0$.   

\Disco{} is second-order accurate in time and space.  It uses the piecewise linear method (PLM) to interpolate the cell-centered primitive values to the cell interfaces for the Riemann fluxes.  The relativistic Harten--Lax--van Leer--Contact (HLLC) approximate Riemann solver \citep{Mignone05} is employed to calculate intercell fluxes, and the time evolution is performed via the second-order total variation diminishing Runge-Kutta (RK2-TVD) algorithm of \cite{Gottlieb98}.  The time step is Courant limited with a typical Courant factor limit of $0.1$.

All simulations in this work are performed in two spatial dimensions ($r$ and $\phi$) using vertically integrated fluid quantities and metric terms evaluated on the equator $z=0$.  We denote the surface density as $\Sig_0 = \int \dd z \rho_0$ and the vertically integrated pressure as $\Pi = \int \dd z P$.

\subsection{The Energy Variable $\tau_U$}
\label{subsec:energy}

Thin accretion disks are highly supersonic, with Mach number $\Mach = u^{\hat{\phi}} \sqrt{1-c_s^2}/c_s \gg 1$, where $c_s = \sqrt{\Gamma P / \rho_0 h}$ is the sound speed.  This is a challenge for hydrodynamics codes, as the specific internal energy $\eps \sim c_s^2$ while the specific kinetic energy $w-1 \sim \tilde{u}^2$, giving $\eps / (w-1) \sim \OO(\Mach^{-2})$.  For hydrodynamics codes written in flux conservative form \eqref{eq:consLaw}, the energy variable must necessarily contain both the kinetic and internal energies.  However, the kinetic energy is due to the bulk motion of the fluid and largely determined by the momentum equations (for Newtonian codes this is exactly true).  The energy equation is solved exclusively to track the internal energy of the fluid. For supersonic flows the internal energy is a small contribution to the total energy.  This makes the internal energy subject to much larger round-off errors than the other fluid quantities, leading to loss of accuracy.

Several schemes exist to combat this issue \citep{Masset00}. The Newtonian \Disco{} code includes the option to specify an exact rotation profile $\Om(r)$, and chooses as its energy variable $\frac{1}{2}\rho v_r^2 + \frac{1}{2}\rho(v_\phi-r\Om)^2 + \rho \eps$; subtracting the kinetic energy associated with $\Om$. This introduces source terms in the energy equation proportional to $\pd_r \Om$, which are exactly known since $\Om(r)$ is exactly specified.  When $\Om$ is chosen close to the fluid $v_\phi / r$ this subtraction allows for accurate evolution of the internal energy even for very thin (high-$\Mach$) disks \citep{Duffell16}.

We choose an energy variable $\tau_U$ that is the relativistic analog to the Newtonian scheme.  Subtracting the kinetic energy associated with some bulk motion can be seen as simply measuring the energy in a particular frame. We specify an exactly known four-velocity $U^\mu(x^\nu)$ chosen to be near the bulk fluid velocity and define the energy as the projection of the stress energy tensor onto this time-like vector $-U_\mu T^{\mu 0}$.  We also perform the standard operation of subtracting the rest-mass energy from the total energy to arrive at our energy variable:
\begin{equation}
	\tau_U = -U_\mu T^{\mu 0} - D \ . \label{eq:tauU}
\end{equation}
This is very similar to the energy variable $\tau$ used by \citep{HARM, Duez05}:
\begin{equation}
	\tau = -n_\mu T^{\mu 0} - D \ , \label{eq:tau}
\end{equation}
where $n^\mu$ is the unit time-like normal vector. In fact, the choice of Equation \eqref{eq:tau} can be seen as just making the choice to measure energy with respect to normal observers.  It is easy to determine: 
\begin{equation}
\tau_U + D = W\left(\tau + D\right) - \gamma^{ij}U_i S_j  \ ,
\end{equation}
where $W = -n_\mu U^\mu$ is the $U$ Lorentz factor in the coordinate frame and $\gamma^{ij}$ is the inverse spatial metric.

If $U^\mu$ is chosen sufficiently close to the fluid velocity, then the dominant component of $\tau_U$ will be the internal energy.  In the case of a thin accretion disk around a black hole, we use a $U^\mu$ that is Keplerian outside the innermost stable circular orbit (ISCO) and smoothly plunging inside.  For a Schwarzschild black hole of mass $M$ this takes the form (in Schwarzschild coordinates)
\begin{align}
	U^0 &= \left \{ \begin{matrix} \frac{2\sqrt{2}/3}{1-2M/r} & r < 6M \\
						\frac{1}{\sqrt{1-3M/r}} & r > 6M \end{matrix} \right . , \nonumber \\
	U^r &= \left \{ \begin{matrix} -\frac{1}{3}\sqrt{\frac{6M}{r}-1} & r < 6M \\
						0 & r > 6M \end{matrix} \right . , \nonumber \\
	U^\phi &= \left \{ \begin{matrix}  \frac{2 \sqrt{3} M}{r^2} & r < 6M \\
						\sqrt{\frac{M/r^3}{1-3M/r}} & r > 6M \end{matrix} \right . . \label{eq:Ugeo}
\end{align}

We find using $\tau_U$ instead of $\tau$ to be essential for accurately evolving thin disks with even moderate Mach numbers.

\subsection{Radiative Cooling}
\label{subsec:cooling}

We restrict our attention to optically thick disks, where radiative cooling occurs at the local blackbody rate. We impose a cooling function \citep{Novikov73, FrankKingRaine}:
\begin{equation}
	\dot{Q} = \frac{8}{3} \frac{\sig_{SB} T^4}{\ka \Sig} , \label{eq:BBcooling}
\end{equation}
where $\sig_{SB}$ is the Stefan--Boltzmann constant, $T = m_p \Pi / \Sig$ is the gas temperature (assuming pure hydrogen), and $\ka $ is the opacity.  We assume that the dominant opacity is due to electron scattering and take $\ka = \ka_{es} = 0.4 cm^2/g$.

The cooling time scale can be much shorter than the local hydrodynamic time scale.  To avoid severe restrictions on the global time step, we use operator splitting to separate the hydrodynamic and cooling evolutions.  Since time evolution in \Disco{} is performed via the method of lines, it is sufficient to prescribe the split scheme to first order in time.

Since the cooling is isotropic to first order in time, it only affects the internal energy of the gas, having no effect on either the surface density or fluid velocity.  As such, for our cooling operator we solve a simple evolution equation for the temperature, leaving $\Sig$ and $u^\mu$ constant.  The change in the conserved variables $\De \mathcal{U}_\text{cool}$ due to the temperature change alone is added to the change from the hydrodynamic evolution.

Schematically, a first-order time step for a single cell begins with primitive variables $\mathcal{P}^i$ and conservative variables $\mathcal{U}^i = \mathcal{U}(\mathcal{P}^i)$.  Over a time step $\De t$ the hydro routines (Riemann fluxes and geometric source terms) add $\De \mathcal{U}^i_\text{hydro}$.  The cooling scheme evolves the initial temperature $T^i$ to a new temperature $T'$ and calculates the change $\De \mathcal{U}^i_\text{cool} = \mathcal{U}(T') - \mathcal{U}^i$, where $\mathcal{U}(T')$ is calculated using the initial values of $\Sig$ and $u^\mu$.  The evolved conservative variables are updated using the sum of the hydro and cooling contributions: 
\begin{equation}
	\mathcal{U}^{i+1} = \mathcal{U}^i + \De \mathcal{U}^i_\text{hydro} + \De \mathcal{U}^i_\text{cool} \ . \label{eq:opsplitU}
\end{equation}
The primitive variables are then calculated accordingly: $\mathcal{P}^{i+1} = \mathcal{P}(\mathcal{U}^{i+1})$.

The temperature evolution equation used to calculate $\De \mathcal{U}_{cool}$, is found from the energy equation obtained by projecting Equation \eqref{eq:GRHD} onto the velocity $u^\mu$.
\begin{equation}
	\Sig u^\mu \nabla_\mu \eps = \Pi \nabla_\mu u^\mu - \dot{Q} \ .
\end{equation}
Neglecting the advection and adiabatic expansion effects of the hydrodynamic evolution during the time step, we are left with a simple equation for the specific internal energy:
\begin{equation}
	\pd_t \eps \approx - \frac{1}{\Sig u^0} \dot{Q} \ .
\end{equation}
Assuming that $\Sig$ and $u^\mu$ are constant during the time step gives the following equation for the temperature evolution due to cooling:
\begin{equation}
	\pd_t T = - \left(\frac{\pd \eps}{\pd T}\right)_{\Sig}^{-1} \frac{\dot{Q}}{\Sig u^0} \ . \label{eq:Tevolution}
\end{equation}
In the simple case of blackbody cooling \eqref{eq:BBcooling} with a constant opacity $\ka_{es}$ and equation of state \eqref{eq:gammalaw} we can integrate Equation \eqref{eq:Tevolution} exactly.  Integrating from $T$ to $T'$ over $\De t$ gives
\begin{equation}
	T' = T \left( 1 + 8 (\Gam-1) \frac{\sig_{SB} T^3}{\ka_{es}\Sig^2 u^0} \De t \right)^{-1/3} \ . \label{eq:Tsolution}
\end{equation}
The adoption of Equation \eqref{eq:opsplitU} with Equation \eqref{eq:Tsolution} makes for an efficient and stable cooling scheme, allowing time steps limited only by the Courant--Friedrichs--Lewy (CFL) condition of the hydrodynamic scheme.

\subsection{Reference Frame and Tidal Forces}
\label{subsec:frameforces}

We perform all simulations in a frame co-orbiting with the secondary black hole.  In this frame the dominant contribution to the metric is that of the secondary black hole itself, which we take to be the Schwarzschild metric of mass $M$ in Kerr--Schild coordinates.  We then perform a coordinate transformation to a frame rigidly rotating with the binary frequency $\Om_\text{bin}$, which is equivalent to adding a shift $\be^\phi = \Om_\text{bin}$ to the metric.  This shift automatically adds both Coriolis and centrifugal forces to the equations of motion for the gas.  The resultant metric is described by the lapse $\al$, shift $\be^i$, and spatial metric $\gam_{ij}$:
\begin{align}
	\al &= \frac{1}{\sqrt{1+2M/r}} \\
	\be^i &= \begin{pmatrix} \frac{2M/r}{1+2M/r} & \Om_\text{bin} \end{pmatrix} \\
	\gam_{ij} &= \begin{pmatrix} 1+2M/r & 0 \\ 0 & r^2 \end{pmatrix}
\end{align}

Incorporating the tidal forces due to the primary cannot be done exactly, as there is no exact metric for an orbiting binary black hole.  We make a pragmatic choice and include the effects of the primary by adding an external force field to Equation \eqref{eq:GRHD}:
\begin{equation}
	\nabla_\mu T^{\mu\nu} = -\dot{Q} u^\nu + f^\nu . 
\end{equation}
This modifies Equation \eqref{eq:sources} as:
\begin{equation}
	\mathcal{S} = \sqrt{-g} \begin{pmatrix} 0 \\
                        \frac{1}{2}T^{\mu\nu}\pd_i g_{\mu\nu} - \dot{Q}u_i  + f_i \\
                        T^{\mu\nu}\nabla_\mu U_\nu + U^\mu u_\mu \dot{Q} - U^\mu f_\mu \end{pmatrix} .\label{eq:sourcesF}
\end{equation}
This prescription, with an appropriately chosen $f^\mu$, captures the main effect of the companion black hole at large binary separation: tidal forces perturbing particle orbits.  Relativistic effects, such as an increased redshift for gas nearer the primary or perturbation of the secondary's horizon, are lost. However, we believe this approximation to be valid when the binary separation is large: $a/M_\text{bin} \gg 1$ and the gravitational field of the primary varies slowly over the domain, or $q \ll 1$, where $q = M_S / M_P$ is the mass ratio and $M_P$ and $M_S$ are the masses of the primary and secondary black holes, respectively.

We calculate the spatial components $f_i$ from the Newtonian potential $\Phi_N$.  In Cartesian coordinates $\vec{x}=(r \cos \phi, r \sin \phi)$ the primary black hole of mass $M_P$ is located at $\vec a = (-a, 0)$.  In the frame of the secondary the potential is
\begin{equation}
	\Phi_N = -\frac{M_P}{|\vec{x} -  \vec{a} |} + \frac{M_P}{a^3}  \vec{a} \cdot \vec{x} \ , \label{eq:phiN}
\end{equation}
where the first term is the gravitational potential of the primary and the second is due to the orbital motion of the origin of the co-orbiting reference frame.  The force $f_i$ is then the gradient of Equation \eqref{eq:phiN}, multiplied by the appropriate energy density:
\begin{align}
	f_r &= \Sig h (u^0)^2 \left( -\cos(\phi) \partial_x  - \sin(\phi)\partial_y\right) \Phi_N \ ,\nonumber \\
	f_\phi &= r\Sig h (u^0)^2 \left( \sin(\phi) \partial_x  -\cos(\phi)\partial_y\right) \Phi_N \ .\nonumber \\
\end{align}
Requiring that the force be orthogonal to the velocity, $u^\mu f_\mu = 0$, then gives $f_0 = -v^i f_i$, completing the prescription of $f_\mu$. This last condition follows from relativistic dynamics and ensures that the source term \eqref{eq:sourcesF} provides no heating or cooling to the gas.


\section{Minidisk Models}
\label{sec:models}

\begin{figure}
\plotone{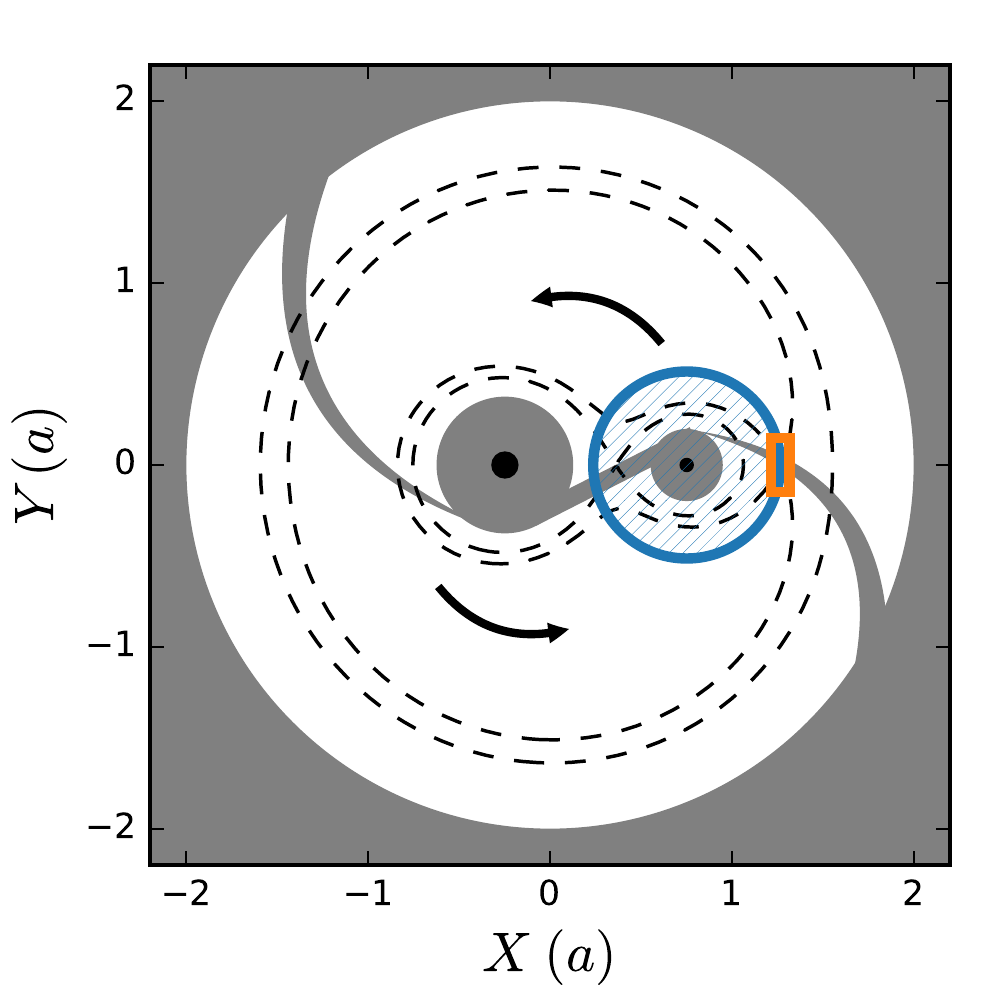}
\caption{\label{fi:domain} Global sketch of circumbinary accretion.  Shown is a binary black hole system (black dots) with orbital separation $a$ surrounded by a circumbinary gaseous disk (gray solid).  A cavity of radius $2a$ is cleared, and streams from the cavity wall feed ``minidisks'' around each black hole. The L1 and L2 Roche curves are plotted as dotted black lines.  The gas streams enter the vicinity of each black hole through the L2 and L3 Lagrange points.  A gaseous ``bridge'' is seen to exchange gas between minidisks in global Newtonian simulations; for simplicity we do not consider it in this work.  The circular computational domain (thick blue circle) is centered around the secondary black hole and extends to the $L2$ Lagrange point.  Arrows denote the rotation direction of the system; the computational domain co-rotates with the binary.  At the boundary near the incoming stream (thick orange rectangle) a nozzle boundary condition with constant mass injection rate $\dot{M}_\text{nozzle}$ is enforced.  Elsewhere the boundary is a diode.} 
\end{figure}

Analysis of circumbinary accretion predicted a cavity to form within $r<2a$, where $a$ is the binary separation \citep{Milos05}. Global Newtonian hydrodynamics simulations confirmed the existence of a cavity but also demonstrated the presence of minidisks around each member of the binary.  These minidisks are fed by streams coming from the cavity wall, revealing the essential role nonaxisymmetry plays in accreting binary systems \citep{Farris14}.  This general picture is sketched in Figure \ref{fi:domain} with the simulation domain.

To model the growth and structure of minidisks, we must specify both the parameters of the binary black hole system and the accretion stream.  The overall mass scale of the binary $M_\text{bin} = M_P+M_S$ sets the overall length scale for the system.  Since we focus on a minidisk around the secondary black hole, we express all lengths in terms of $M = M_S$.  Since we work in units where $c=1$, this also sets the fundamental time scale of the system. The use of physical constants in the cooling prescription \eqref{eq:BBcooling} introduces a mass scale $\bar{m}=$ into the system, which scales with $M$ as $\bar{m} \propto M^{5/2}$.  

Our implementation of frame and tidal forces (see Section \ref{subsec:frameforces}) restricts our analysis to large binary separations $a$ and small mass ratios $q$.  We fix $q=0.11$ and $a = 100 M_\text{bin} \approx 1000 M$ for all runs, which satisfies these requirements.  At this separation the corrections to our tidal force prescription are at most $\OO({M_P / a}) \sim 1\%$, but the orbital time scale is still in an accessible regime.  The orbital angular velocity is $\Om_\text{bin} = \sqrt{M_\text{bin} / a^3} \approx 10^{-4} M^{-1}$ and the orbital period is $T_\text{bin} = 2\pi / \Om_\text{bin} \approx 2 \pi \times 10^4 M$.

A circular binary emits GWs that carry away energy and angular momentum and eventually lead to merger.  The time for a circular binary to merge \citep{Peters64}, in units of the initial orbital period, is
\begin{equation}
	T_\text{merge} / T_\text{bin} = \frac{5}{512 \pi}\frac{(1+q)^2}{q} \left(\frac{a}{M_\text{bin}}\right)^{5/2} \ . \label{eq:Tmerge}
\end{equation}
For the system we consider $a=100 M_\text{bin}$ and $q=0.1$, so $T_\text{merger} \approx 3.7\times 10^3 T_\text{bin}$.  The simulations run for $\approx 30 T_\text{bin}$, so we do not include evolution of the binary orbital parameters at this time.

We model the accretion streams as radial infall through the L2 Lagrange point of the binary, based on global Newtonian simulations of circumbinary accretion \citep{Farris14, Farris15A, Farris15B, DOrazio12, DOrazio16}.  The radial velocity is set to be $v^r = -1/2 v_\text{bin}$, where $v_\text{bin} = \sqrt{M_\text{bin}/a} \approx 0.1$ is the binary orbital velocity.  The angular velocity of the stream is zero in the co-rotating frame.  Since the streams are ballistic, the sound speed should be significantly less than the stream velocity.  To ensure this, we set the pressure in the stream as $\Pi = 5 \times 10^{-6} \Sig$.  Several values of this parameter were used during testing; we found that they did not affect the resulting minidisk. Rather, shock heating and radiative cooling allow the gas to find its own equilibrium temperature once it is incorporated into the disk.  The stream is given a width of $ \De \phi = 0.4$ rad, approximately that of the streams seen in global Newtonian simulations \cite{Farris14}.

The density in the nozzle is set by the accretion rate $\dot{M}$, the main parameter of interest in this study.  The inclusion of dimensionfull $\ka$ and $\sig_{SB}$ parameters in the cooling term breaks the scale invariance of mass energy that would otherwise be present.  Streams of different $\dot{M}$ will cool at different rates relative to the orbital period $T_{\text{bin}}$, leading to hotter or cooler disks.  The density in the stream has a profile $\Sig \propto \cos^2(\pi \phi / \De \phi)$, with the normalization set to match the total accretion rate of specified $\dot{M}$.

The numerical grid is centered on the secondary black hole and extends from $r_\text{in} = 4 M$ to $R_\text{L2} \approx 358 M$, the radius of the L2 Lagrange point. Radial zones are distributed logarithmically, and azimuthal zones are placed to keep the aspect ratio of cells close to unity. The stream extends over $\phi \in [-\De \phi / 2, \De \phi / 2] rad$, boundary cells within the stream are fixed to their local stream values. On the outer boundary away from the stream a diode boundary condition is used: zero gradient in all fluid variables with the radial velocity restricted to be positive or zero.

At the inner boundary a hybrid boundary condition is used.  The fluid velocity is set to be exactly as given in Equation \eqref{eq:Ugeo}, appropriate for ballistic matter infalling on geodesics.  The density is set to be $\Sig = -\dot{M} / r U^r \De \phi_\text{cell}$, where $\dot{M}$ is calculated from the innermost non-boundary annulus and $\De \phi_\text{cell}$ is the angular width of the cell.  Given $\Sig$, the pressure $\Pi$ is set to ensure isentropic infall.

Although the inner boundary is outside the event horizon, we find that it does not affect the evolution of the system.  This is because it is still inside the sonic radius of the flow, so no information can propagate out to the minidisk itself.  See the discussion in Section \ref{subsec:bc} and Figure \ref{fi:bc_hr_comp} for details. The CFL limited time step $\De t$ is controlled by the innermost zones of the grid, where the radial velocity is large (due to the black hole) and the cells are narrow.  Placing the inner boundary above the event horizon allows for much larger time steps than otherwise possible; keeping it below the sonic point ensures the fidelity of the simulation.  We found $r_{in} = 4M$ to be a good choice.

The initial condition for each minidisk is an $\al=10^{-3}$ Novikov--Thorne accretion disk with $\dot{M}$ set equal to the nozzle rate \citep{Novikov73}.  

\begin{deluxetable}{ccccccc}
\tablecaption{Minidisk Models \label{tb:models}}
\tablehead{\colhead{Name} &\colhead{$\dot{M}$\tablenotemark{a}} & \colhead{$r_\text{in}$} & \colhead{$q$} & \colhead{$N_r$} & \colhead{$T_\text{start}$\tablenotemark{b}} & \colhead{$T_\text{end}$\tablenotemark{b}}}
\startdata
\model{1} & $1.9\times10^{3}$  & $4M$ & $0.11$ & $256$ & $0$ & $29$ \\
\model{1.5} & $5.8\times10^{2}$ & $4M$ & $0.11$ & $256$ & $0$ & $29$ \\
\model{2} & $1.9\times10^{2}$ &  $4M$ & $0.11$ & $256$ & $0$ & $29$ \\
\model{2.5} & $5.8\times10^{1}$ & $4M$ & $0.11$ & $256$ & $0$ & $25$ \\
\model{3} & $1.9\times10^{1}$ & $4M$ & $0.11$ & $256$ & $0$ & $29$ \\
\model{2-hr} & $1.9\times10^{2}$ & $4M$ & $0.11$ & $512$ & $28$ & $32$ \\
\model{2-bc} & $1.9\times10^{2}$ & $1.8M$ & $0.11$ & $256$ & $25$ & $26$ 
\enddata
\tablenotetext{a}{$\dot{M}$ is given in code units and scales as $M^{3/2}$.}
\tablenotetext{b}{$T_\text{start}$ and $T_\text{end}$ are given in units of $T_\text{bin}$.}
\end{deluxetable}

The five primary minidisk models in this study are summarized in Table \ref{tb:models}.  The accretion rate is given in code units, which scale as $M^{3/2}$.  If $M=M_{\odot}$, the accretion rate for \model{2} corresponds to $2\times10^{-3} M_{\odot} \text{yr}^{-1}$.  At this scale $T_\text{bin} = 0.3$ s and $T_\text{merge} = 20$ minutes.  For an SMBHB with $M_\text{bin} = 10^6 M_{\odot}$, $T_\text{bin} = 8.6$ hr and $T_\text{merge} = 3.7$ yr.  Two tests, Models \texttt{2-hr} and \texttt{2-bc}, were run to check dependence on numerical resolution and the inner boundary condition, respectively. Due to resource constraints, \model{2-bc} was only run for a single orbit, and \model{2-hr} for four orbits.

\subsection{Fiducial Run}
\label{subsec:fiducial}

\begin{figure}
\plotone{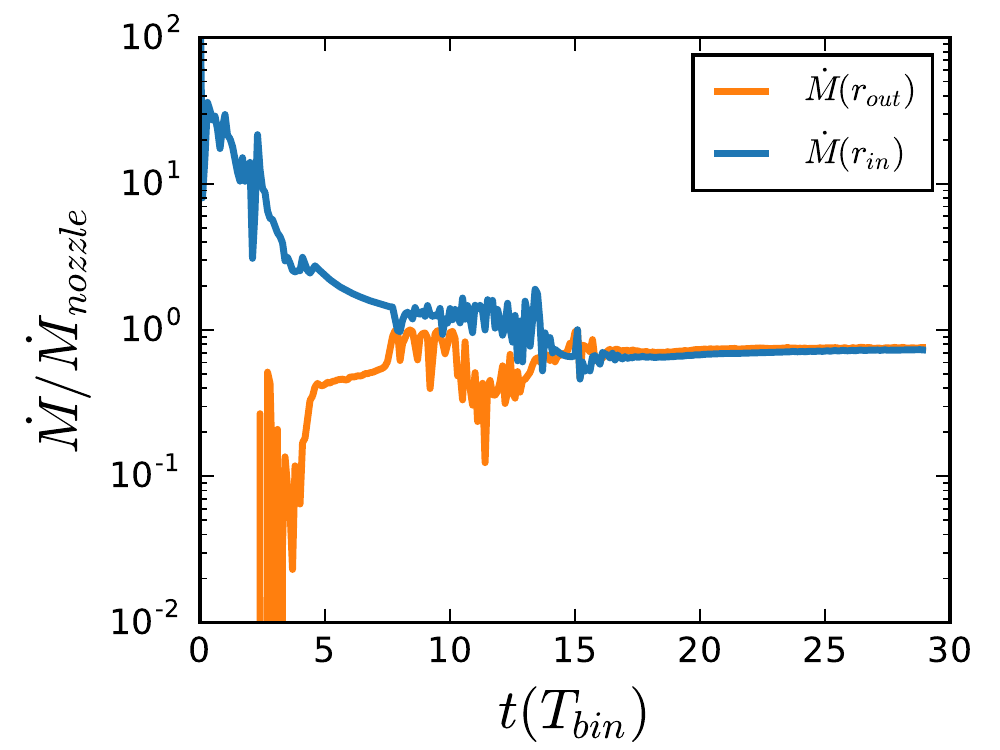}
\caption{\label{fi:mdot-fid} Time series of the accretion rate (measured as a fraction of the nozzle mass injection rate) through the inner (blue) and outer (orange) boundaries of \model{2}.  After $\sim17$ orbits the inner and outer accretion rates balance, indicating the onset of a quasi-steady state.}
\end{figure}

\begin{figure}
\plotone{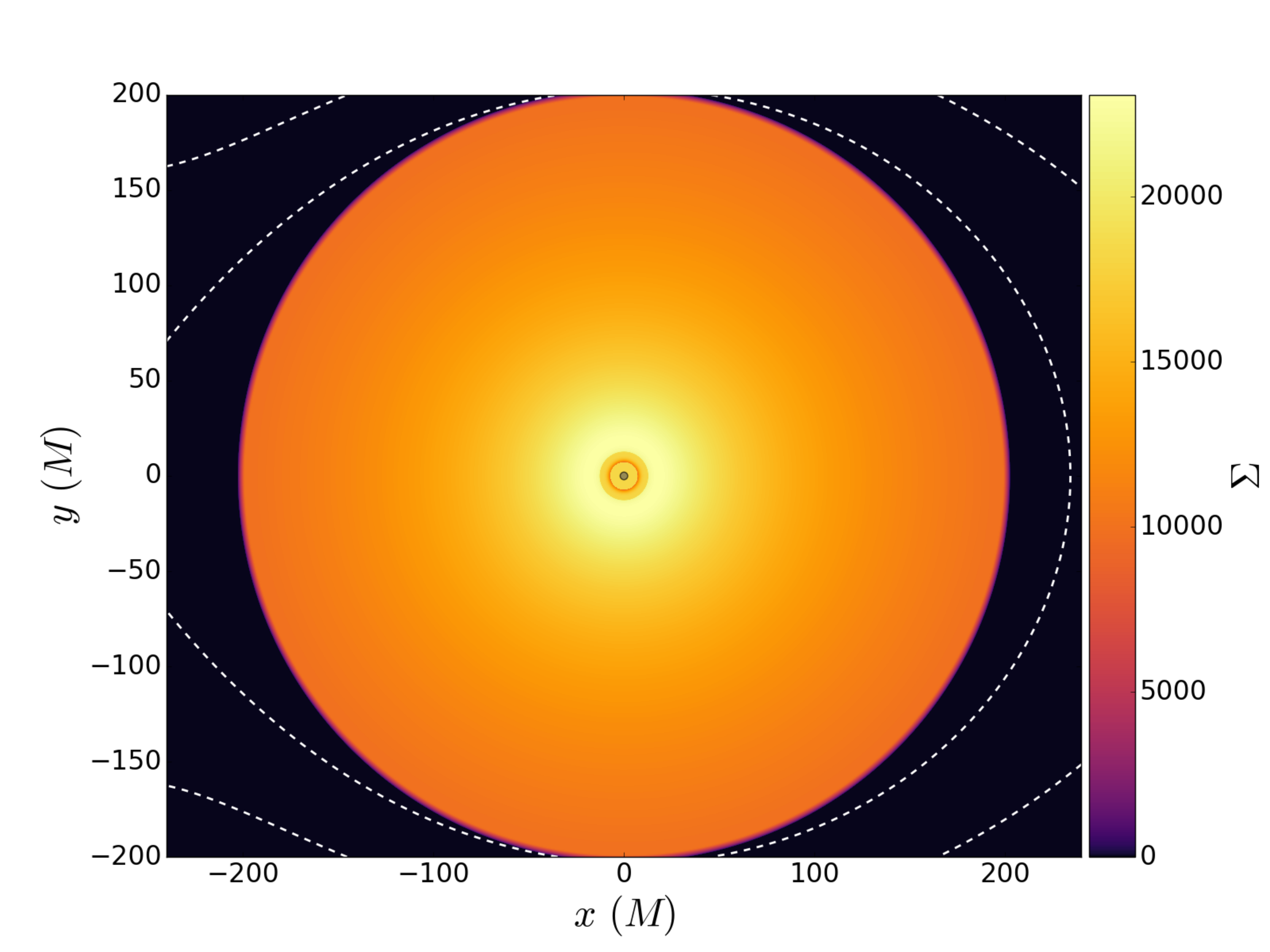}
\caption{\label{fi:sig-fid-0} Initial surface density for fiducial minidisk.  Magenta dashed lines are level curves of the Roche potential corresponding to the L1 and L2 Lagrange points.}
\end{figure}

\begin{figure}
\plotone{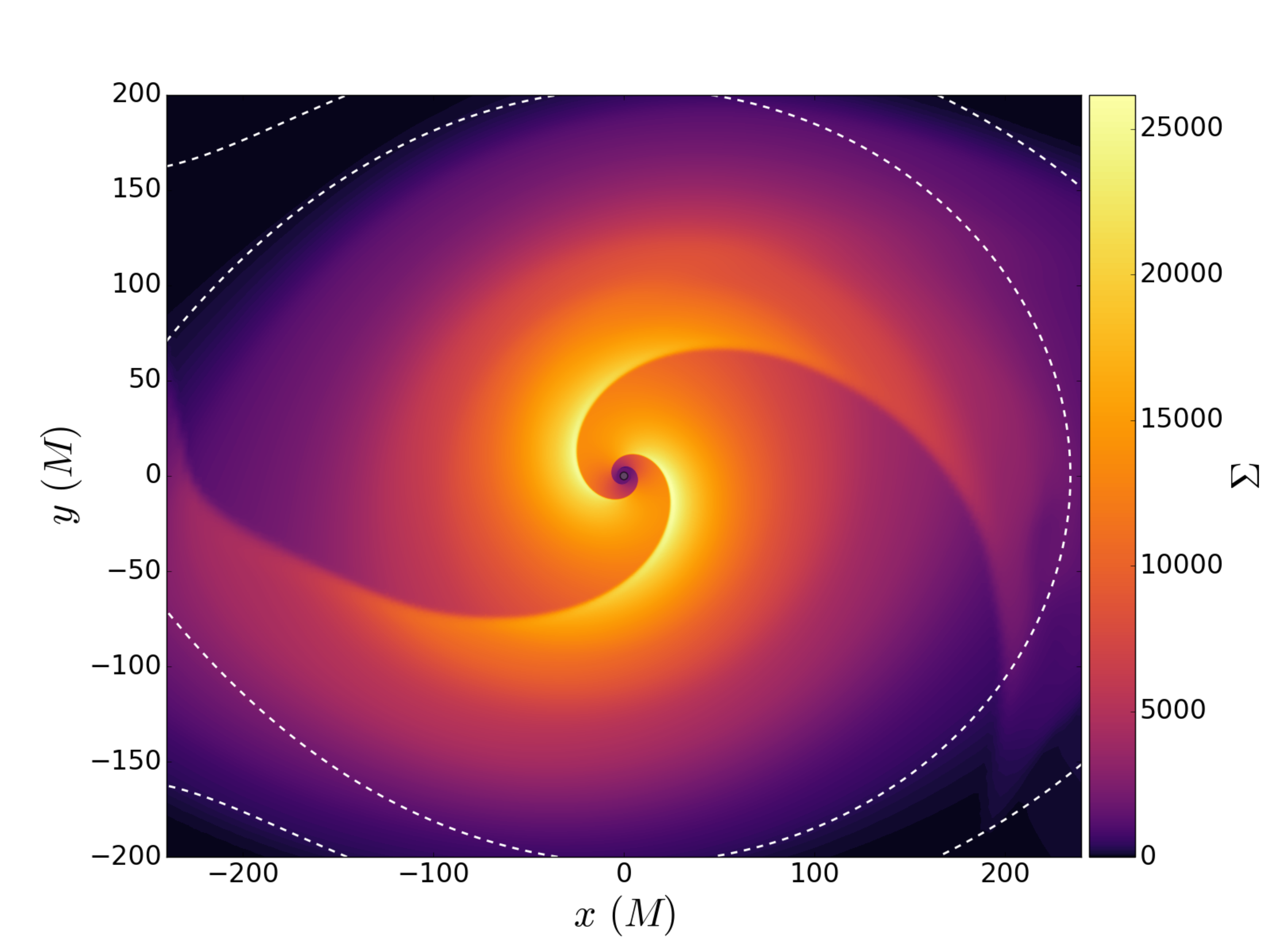}
\caption{\label{fi:sig-fid-05} Same as Figure \ref{fi:sig-fid-0}, but at $t = 1/2\ T_\text{bin}$.  The initial disk quickly develops spiral shocks.}
\end{figure}

\begin{figure}
\plotone{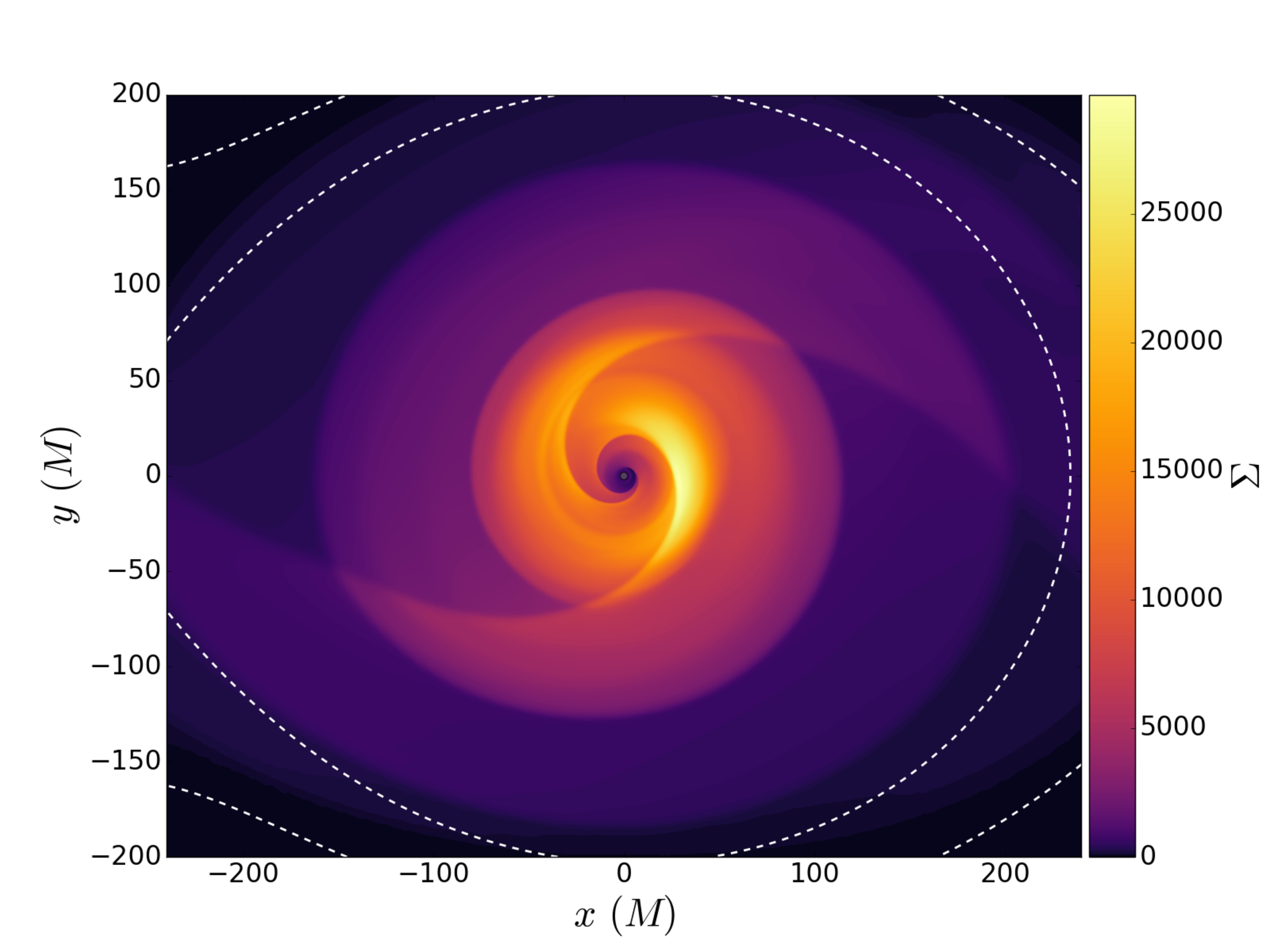}
\caption{\label{fi:sig-fid-2} Same as Figure \ref{fi:sig-fid-0}, but at $t = 2 T_\text{bin}$.  Accretion proceeds in clumps as gas from the initial disk is either thrown out of the domain or falls into the black hole.}
\end{figure}

\begin{figure}
\plotone{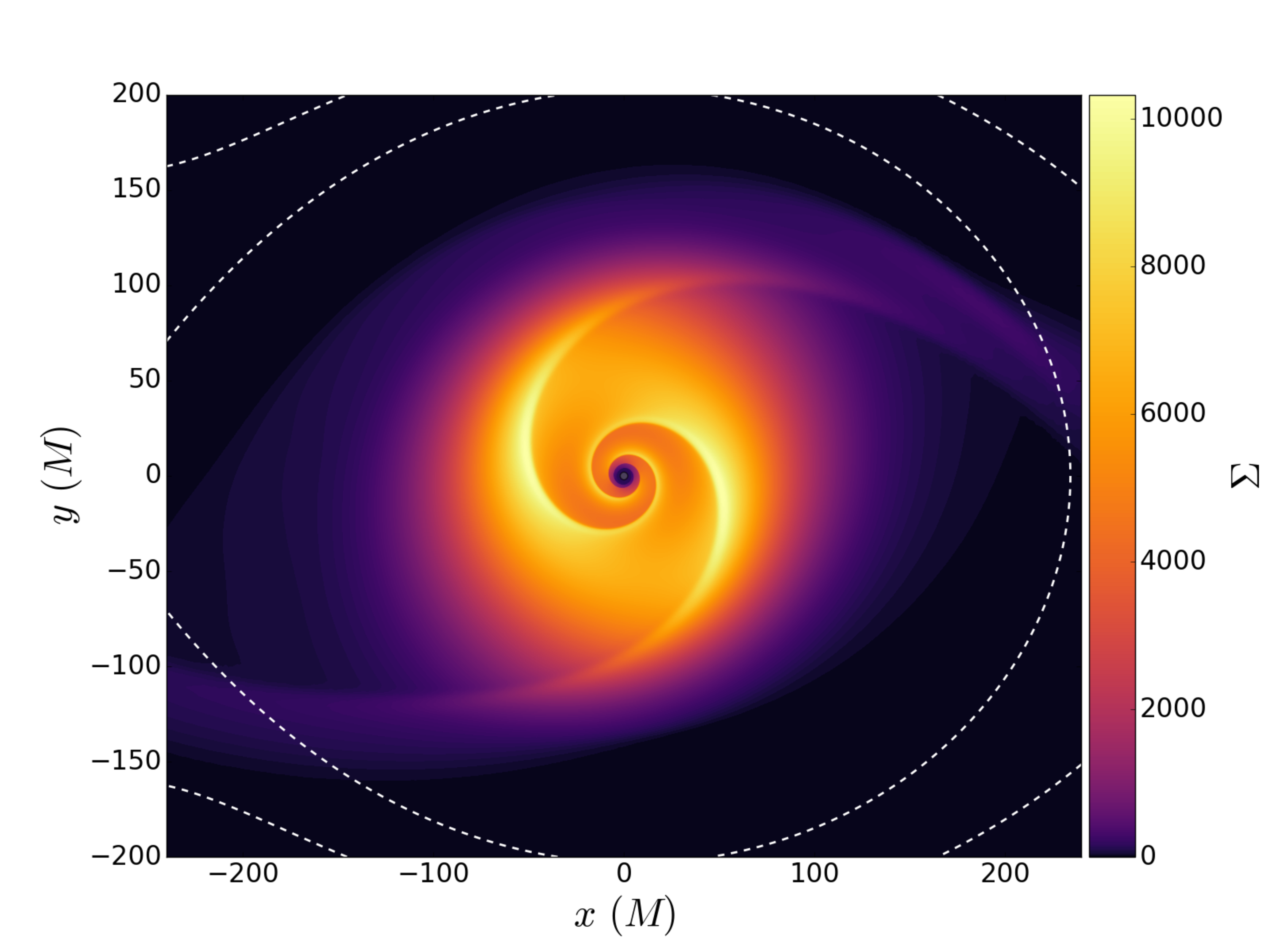}
\caption{\label{fi:sig-fid-5} Same as Figure \ref{fi:sig-fid-0}, but at $t = 5 T_\text{bin}$.  Most of the initial gas has left. The accretion rate drops as the disk cools.}
\end{figure}

We take the $\dot{M} = 2 \times 10^{-3} M_\odot \text{yr}^{-1}$ model (\model{2}) as our fiducial run.  The accretion rate through the inner ($r=4 M$) and outer ($r=r_\text{L2}$) boundaries is plotted as a function of time in Figure \ref{fi:mdot-fid}. The initial disk has a surface density much higher than the stream from L2 (Figure \ref{fi:sig-fid-0}).  In the first orbit the initial disk develops its own tightly wound two-armed spiral as the gas in the outer radii is flung away.  The two-armed spiral begins tightly wound and diffuse, but quickly sharpens into a shock and begins to open as the disk heats (Figure \ref{fi:sig-fid-05}).  Some gas flung to the boundary falls back, and accretion proceeds in clumps for the first two orbits (Figure \ref{fi:sig-fid-2}).  After two orbits, the bulk of the initial disk has either been accreted or thrown out the outer boundary.  As the remaining disk accretes, it becomes more diffuse and cools, the spiral shocks tighten, and the accretion rate drops (Figure \ref{fi:sig-fid-5}).

\begin{figure}
\plotone{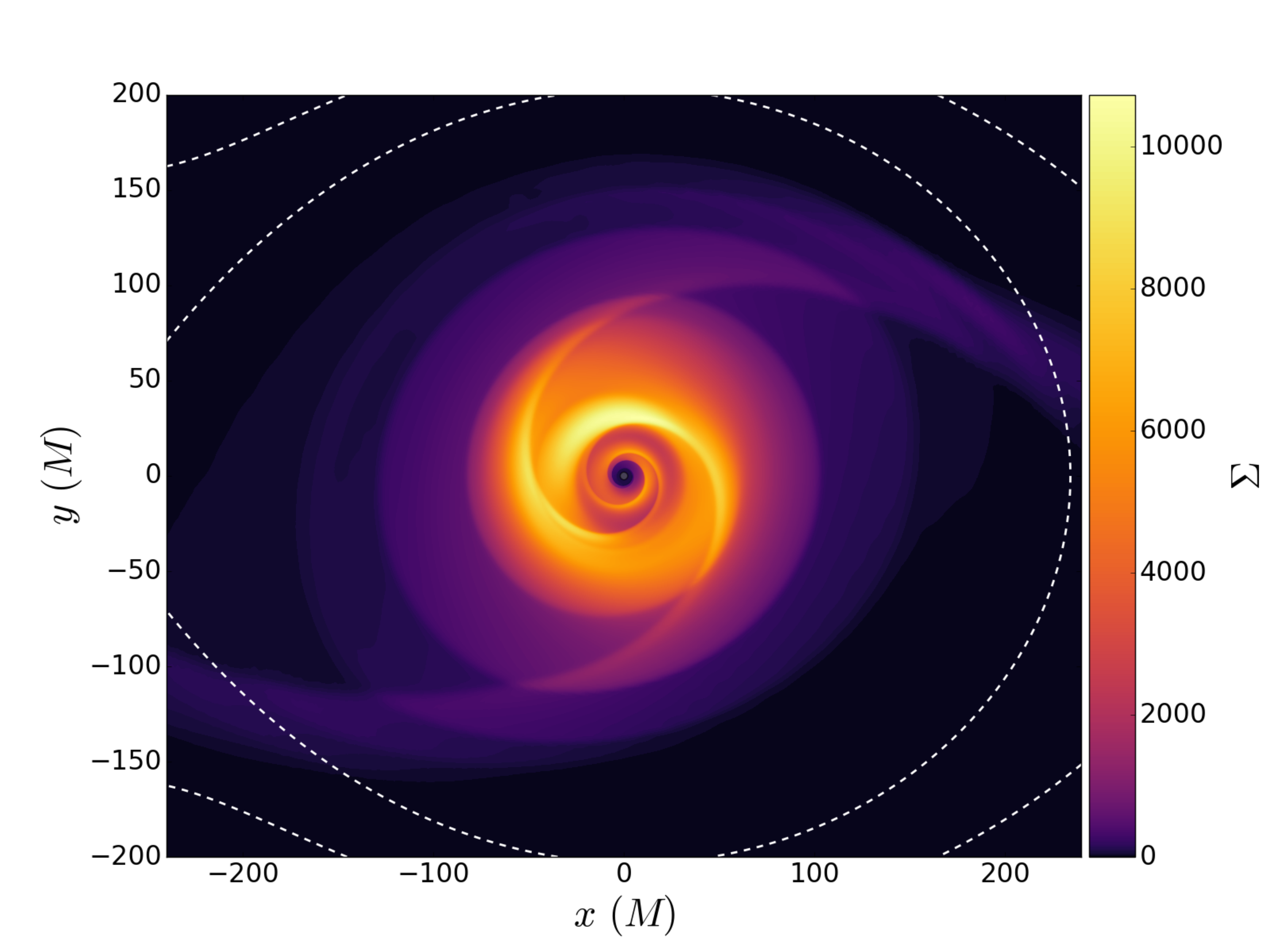}
\caption{\label{fi:sig-fid-13} Same as Figure \ref{fi:sig-fid-0}, but at $t = 13 T_\text{bin}$.  Variability sets in as the disk begins interacting dynamically with the accretion stream.}
\end{figure}

At $t\sim 8 T_\text{bin}$ the accretion rate through the disk becomes comparable to the stream.  For the next eight orbits a complicated and turbulent interaction occurs between the disk and stream.  The stream begins wobbling and sending gas to the disk in clumps.  Clumps execute an orbit of the black hole and collide with the original stream, causing it to wobble further, producing more clumps.  The clumps are sheared by the accretion flow, and after many orbits within the disk, they are accreted. The spiral shocks retain their structure through the clumpy accretion (Figure \ref{fi:sig-fid-13}).

\begin{figure*}
\plotone{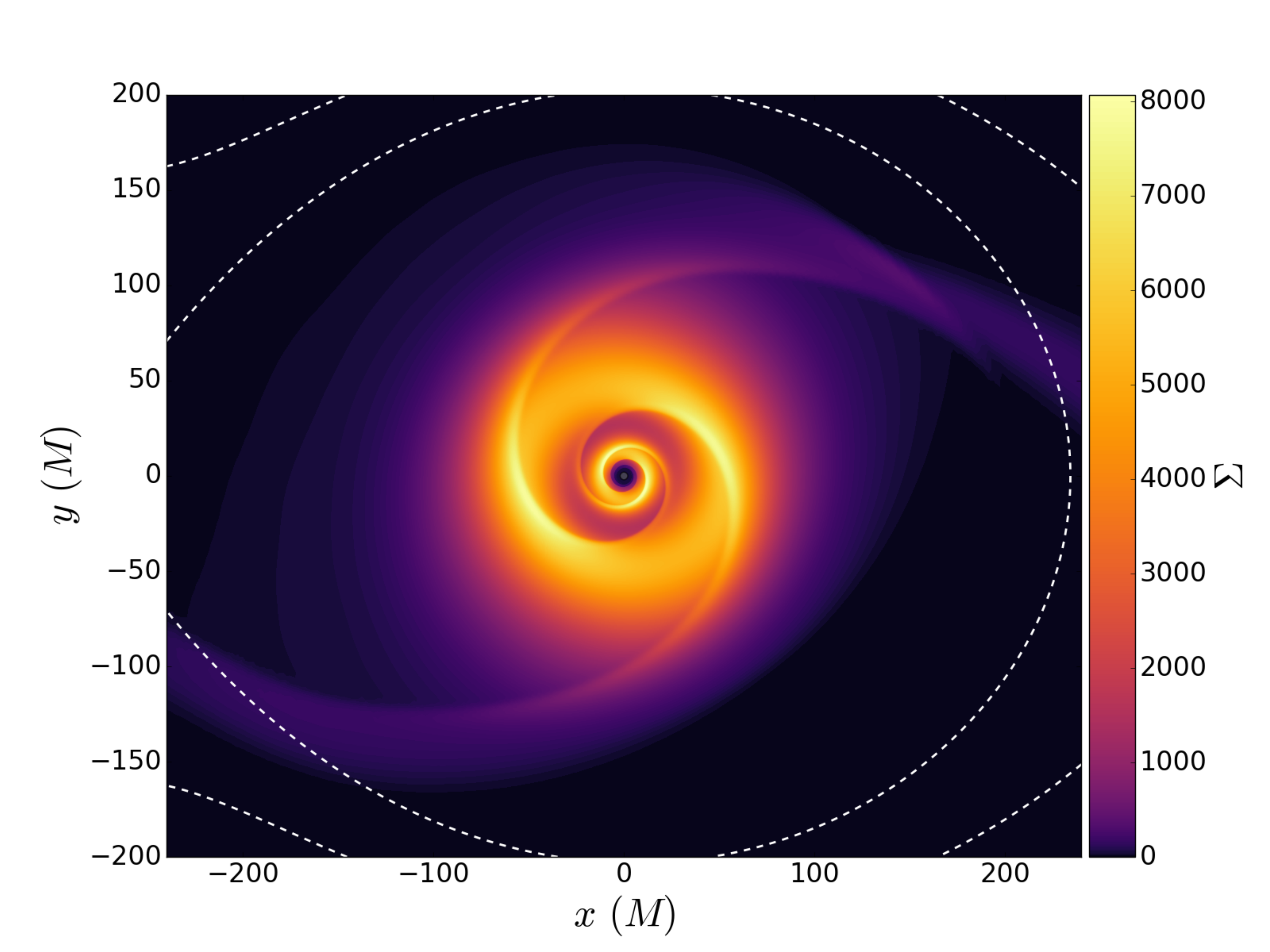}
\caption{\label{fi:sig-fid-28} Same as Figure \ref{fi:sig-fid-0}, but at $t = 28 T_\text{bin}$.  The disk in a quasi-steady state.}
\end{figure*}

At $t\sim17 T_\text{bin}$ the disk--stream interaction stabilizes and the disk reaches a quasi-steady state.  The accretion rate through the inner boundary matches the inflow rate through the outer boundary, and the two-armed spiral shocks lock into a constant pattern.  Small-scale flows still occur as gas slowly redistributes within the disk, relaxing to a true steady state.  The simulation ends at $T\sim30 T_\text{bin}$ (Figure \ref{fi:sig-fid-28}), before this secular evolution ends.

\section{Analysis}
\label{sec:analysis}

\subsection{Shock Detection}
\label{subsec:shockDet}

To characterize the role of spiral shocks, we first must find them in the simulation output.  We use a modified version of the relativistic shock detector presented in \cite{Zanotti10}.  The basic algorithm classifies adjacent zones (e.g. in the $\phi$ direction) by considering the Riemann problem at their interface.  Interfaces are classified as shocks if the Riemann problem solution requires at least one shock.  We find this to be too sensitive a criterion, so we add additional constraints for an interface to be considered a shock.

In the simple algorithm, first the relative normal velocity $v_{12}$ is calculated.  Assuming $\phi$-separated cells $1$ and $2$ with $\Pi_1 > \Pi_2$, this is
\begin{equation}
	v_{12} = \frac{v^{\hat{\phi}}_1 - v^{\hat{\phi}}_2}{1 - v^{\hat{\phi}}_1 v^{\hat{\phi}}_2} \ . \label{eq:vrel}
\end{equation}
In the above $v^{\hat{\phi}}$ is the azimuthal 3-velocity in an orthonormal frame. The relative velocity $v_{12}$ is compared to the critical velocity $(\tilde{v}_{12})_{\mathcal{R} \mathcal{S}}$:
\begin{equation}
	(v_{12})_{\mathcal{R}\mathcal{S}} = \tanh \int_{\Pi_1}^{\Pi_2} \frac{\sqrt{h(\Pi)^2 + \mathcal{A}_1^2(1-c_s(\Pi)^2)}}{(h(\Pi)^2 + \mathcal{A}_1^2)\Sig(\Pi) c_s(\Pi)} \dd \Pi \ . \label{eq:vrs}
\end{equation}  
In the integral in Equation \eqref{eq:vrs}, thermodynamic quantities $h$, $c_s$, and $\Sig$ are calculated at the specific entropy $s_1$ of cell 1 and $\mathcal{A}_1 = h_1 u_1^{\hat{r}}$.  This velocity is the largest normal velocity that a single rarefaction wave can provide between cells 1 and 2    \citep{Rezzolla03}.  If $v_{12} > (\tilde{v}_{12})_{\mathcal{R}\mathcal{S}}$ a shock is necessarily present.

We find the criterion $v_{12} > (\tilde{v}_{12})_{\mathcal{R}\mathcal{S}}$ alone to be far too sensitive, so to consider an interface a shock, we require $v_{12} - (\tilde{v}_{12})_{\mathcal{R}\mathcal{S}} > \Delta v_{amb}$, where $\Delta v_{amb}$ is a threshold value.  Lastly, we require the specific entropy to increase in the direction of the flow: $u^\phi \partial_\phi s > 0$.

Our full shock detector algorithm runs on each annulus of the computational grid.  Since the shocks tend to have small pitch angles, they have small radial width but may extend over a few zones azimuthally.  The algorithm identifies all zones in the annulus separated by a shock via our criteria above and groups them into contiguous segments.  The two segments with the largest values of $v_{12} -  (\tilde{v}_{12})_{\mathcal{R}\mathcal{S}} $ are classified as the shocks. The threshold $\Delta v_{amb}$ is set for each annulus to be the larger of $0$ or the fourth-largest maxima of $v_{12} -  (\tilde{v}_{12})_{\mathcal{R}\mathcal{S}} $. We find this value to have good discriminatory power for these runs, picking out the shocks but ignoring the ambient flow.

The final output of the shock detector is the position $\Phi(r)$ for the leading and trailing edge of both shock waves at each annulus of the computational grid.

\subsection{Wave Propagation}

A linear perturbation to a fluid quantity of azimuthal mode $m$ and pattern speed $\Omega_P$ can be written as
\begin{equation}
	\delta X(t, r, \phi) = \tilde{X}_{m,\omega}(r) \exp \left( i \int_r k(r) \dd r  + i m (\phi - \Omega_P t) \right)\ . \label{eq:pert}
\end{equation}
Such perturbations are spirals with pitch angle
\begin{equation}
	\tan \theta = m / k r \ . \label{eq:pitch}
\end{equation}

The hydrodynamics equations for such a mode reduce to a system of ordinary differential equations in the radial coordinate $r$.  In the `tight winding' limit $k \gg \tilde{X}'/\tilde{X}$, the perturbation is highly oscillatory, and the WKB approximation may be used to yield the well-known \citep{Binney08} dispersion relation for tightly wound waves in a Newtonian gaseous disk:

\begin{equation}
	\Omega^2 - m^2(\Omega-\Omega_P)^2 + c_s^2 k^2 = 0 \ . \label{eq:dispNewt}
\end{equation}
The generalization of Equation \eqref{eq:dispNewt} to a disk in the Schwarzschild metric is straightforward \citep{Perez97}.  The background state is a steady thin disk of sound speed $c_s \ll 1$ and velocity given by Equation \eqref{eq:Ugeo} in the region $r > r_{ISCO}$.  Assuming that the perturbation is adiabatic, tightly wound, and neglecting vertical structure leads to the dispersion relation for relativistic $p$-modes \citep{Abramowicz13}:
\begin{equation}
	\left(1-\frac{6M}{r} \right)\left(U^\phi\right)^2 - m^2(U^\phi-U^0\Omega_P)^2 + g_{rr} c_s^2 k^2 = 0 \ . \label{eq:disprel}
\end{equation}
In Schwarzschild coordinates the perturbation remains a spiral with pitch angle \eqref{eq:pitch}.  The dispersion relation can be solved for the radial wavenumber $k$ yielding
\begin{align}
	\tan \theta &= \left( \frac{r U^\phi}{c_s} \right)^{-1} \left(1-\frac{2M}{r}\right)^{\frac{1}{2}} \label{eq:dispRelPitch} \\ \nonumber
		&\times \left(\left(1- \frac{U^0 \Omega_P}{U^\phi}\right)^2 - \frac{1}{m^2}\left(1-\frac{6M}{r}\right)\right)^{-\frac{1}{2}}\ . 
\end{align}
Shocks, of course, are intrinsically nonlinear perturbations to a fluid flow.  Weak shocks, however, travel very near the local sound speed and can be approximated as linear waves.  The degree to which spiral shocks in the minidisks satisfy Equation \eqref{eq:dispRelPitch} can be used to gauge their nonlinearity.  

\subsection{Angular Momentum Decomposition}
\label{subsec:angMomDecom}

A detailed look at the angular momentum transport due to spiral shocks requires a decomposition of the various torques acting on the system. Defining angular integrated quantities as $\avet{\cdot} = \int \dd \phi \sqrt{-g} (\cdot)$, we can write effective 1D continuity and angular momentum equations \eqref{eq:GRHD} in a coordinate frame:
\begin{align}
	\partial_t \ave{\Sig u^0} + \partial_r \ave{\Sig u^r} &= 0 \ ,\label{eq:aveAng}\\
	\partial_t \ave{\Sig h u^0 u_\phi} + \partial_r \ave{\Sig h u^r u_\phi} &= \ave{f_\phi}- \ave{u_\phi \dot{Q}_{cool}} \ .\nonumber
\end{align} 
We can decompose Equation \eqref{eq:aveAng} by separating the part that strictly obeys the continuity equation.  First, define
\begin{align*}
	\aveRe{\Sig h u^\mu u_\phi} &= \ave{\Sig h u^\mu u_\phi} - \ave{\Sig u^\mu} \ell \ , \\
	\ell &= \frac{\ave{\Sig h u^0 u_\phi}}{\ave{\Sig u^0} } \ .
\end{align*}
We can then write angular momentum conservation as
\begin{equation}
	\ave{\Sig u^\mu} \partial_\mu \ell + \partial_\mu \aveRe{\Sig h u^\mu u_\phi} =  \ave{f_\phi}- \langle u_\phi \dot{Q}_\text{cool} \rangle \ , 
\end{equation}
or
\begin{align}
	\ave{\Sig u^0} \partial_t \ell &= \dot{M} \partial_r \ell -  \partial_r \aveRe{\Sig h u^r u_\phi} +  \ave{f_\phi}- \langle u_\phi \dot{Q}_\text{cool} \rangle \ , \nonumber \\
	&\equiv \tau_{\dot{M}} + \tau_\text{Re} + \tau_\text{ext} + \tau_\text{cool} \ . \label{eq:angMomDecom}
\end{align} 
Equation \eqref{eq:angMomDecom} decomposes the rate of change of angular momentum into four contributions: accretion ($\tau_{\dot{M}}$), Reynolds-type stress ($\tau_\text{Re}$), external torques ($\tau_\text{ext}$), and cooling ($\tau_\text{cool}$).  Accretion torque is due to the bulk flow of gas over a radially varying specific angular momentum profile. The Reynolds torque is due to nonaxisymmetric structures in the flow, particularly varying azimuthal profiles of radial mass flux and specific angular momentum. External torques in our case are completely due to the presence of the binary companion. The cooling torque $\tau_\text{cool}$ is a purely relativistic effect, representing the loss of momentum to photons. It is small so long as the rest-mass energy density is greater than the thermal energy of the gas.

\subsection{Ray-traced Spectra}
\label{subsec:raySpec}

Using the physical cooling prescription \eqref{eq:BBcooling} gives us the ability to self-consistently calculate the electromagnetic emission of our minidisk models.  Since the bulk of the emission comes from the innermost regions of the disk, a radiative transfer simulation in the curved spacetime of the black hole is necessary to accurately produce an observational spectrum.

We follow the method of \cite{Kulkarni11} and \cite{Zhu12} to perform the radiative transfer.  We set up an image plane a large distance $d = 10^6 M$ from the central black hole.  The flux through the image plane is
\begin{equation}
	F_\nu = \frac{1}{d^2} \int \dd A I_\nu \ ,
\end{equation}
where $I_\nu$ is the specific intensity at the image plane.  Null rays $k^\mu$ are integrated from the image plane through the Schwarzschild spacetime to the disk surface at $z=0$.  The effects of finite disk height were investigated in \cite{Kulkarni11}, and found to be negligible for moderate observer inclination angles $i$. Assuming vacuum between the image plane and disk, $I_\nu / \nu^3$ is conserved along each ray.  We can then write
\begin{equation}
	F_\nu = \frac{1}{d^2} \int \dd A \left( \frac{\nu}{\nu'} \right)^3 I_{\nu'}' \ ,
\end{equation}
where $\nu'$ and $I_{\nu'}'$ are the frequency and specific intensity of the ray, respectively, as measured in the comoving fluid frame of the disk. In our cooling model \eqref{eq:BBcooling} the radiated emission is blackbody and isotropic in the rest frame of the gas.  Hence,
\begin{equation}
	I_{\nu'}' = B_{\nu '} (T_\text{eff}) = \frac{2 {\nu'}^3}{\exp\left(\nu' / k_B T_\text{eff}\right) - 1}\ ,
\end{equation}
where
\begin{equation}
	T_\text{eff} = \left(\frac{\dot{Q}_\text{cool}}{2 \sig_{SB}} \right)^{1/4} \ .
\end{equation}
Altogether,
\begin{equation}
	F_{\nu} = \frac{1}{d^2} \int \dd A \frac{2 \nu^3}{\exp\left(\nu' / k_B T_\text{eff}\right) - 1} \ .
\end{equation}
The frequency $\nu'$ is obtained from the null ray $k^\mu$.  With $u^\mu$ the velocity of the disk, $u_{im}^\mu$ the velocity of the static image plane, and $k^\mu(x_{im})$ and $k^\mu(z=0)$ the values of $k^\mu$ at the image plane and disk, respectively:
\begin{equation}
	\nu' = \frac{u^\mu k_\mu(z=0)}{u^\mu_{im} k_\mu(x_{im})} \ \nu \ .
\end{equation}
The image plane is an elliptical grid of points laid out in a similar manner to \cite{Kulkarni11} at $d=10^6M$ and inclination $i = 0^\circ, 15^\circ, 30^\circ, 45^\circ, 60^\circ, 75^\circ$ arranged normal to the radial direction.  In Cartesian coordinates local to the image plane the grid points are located at $x_{jk} = b_j \cos \phi_k$ and $y_{jk} = b_j \cos i \sin \phi_k$.  The points $b_j$ are logarithmically distributed between $2M$ and $400M$, and the  $\phi_k$ are distributed linearly in $[0, 2\pi)$. The effective temperature $T_\text{eff}$ and ratio $\nu' / \nu$ are calculated at each ray's origin point in the disk.  Rays that lie outside the simulation domain are ignored and set to zero specific intensity.

We do not consider the effects of the primary black hole and the orbital motion of the simulation domain.  The effects of the former should be small given the orbital separation $a = 100M_\text{bin}$, and the effect of the latter would be a orbital phase dependent doppler beaming on the order of $v_\text{bin}$.


\section{Results}
\label{sec:results}

\begin{figure}
\plotone{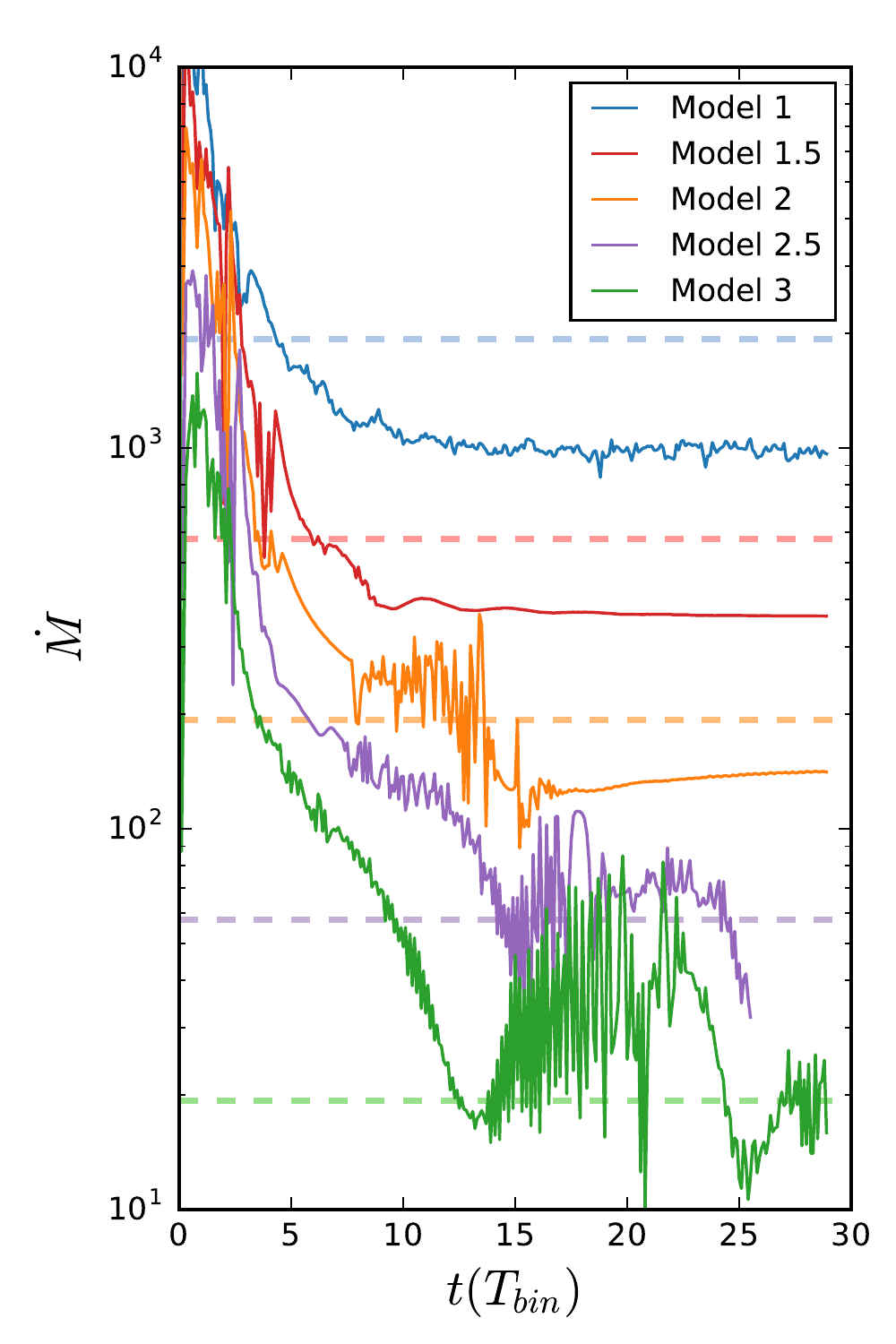}
\caption{\label{fi:mdot_all} Accretion rate through the inner boundary as a function of time for \model{1} (blue), \model{1.5} (red), \model{2} (orange), \model{2.5} (purple), and \model{3} (green).  $\dot{M}_\text{nozzle}$ for each model is shown as a dashed line in the corresponding color. Variability lasts longer and is of greater amplitude in disks with lower accretion rates.  Models \texttt{1}, \texttt{1.5}, and \texttt{2} establish a quasi-steady state.}
\end{figure}

The accretion rate through the inner boundary as a function of time is plotted in Figure \ref{fi:mdot_all} for all primary models.  In the first few orbits the accretion rate is very large, as the initial disk is disrupted by the tidal potential and incoming accretion stream.  This is typically followed by a phase of highly variable behavior before accretion settles into a quasi-steady state.  Models \texttt{1} and \texttt{1.5} reach their asymptotic accretion rates in the first dozen orbits, \model{2} after 17 orbits following a dynamic phase described in Section \ref{subsec:fiducial}, and Models \texttt{2.5} and \texttt{3} remain highly variable for the duration of the simulation.  The asymptotic values of the accretion rate into the black hole are always smaller than the injection rate $\dot{M}_\text{nozzle}$, as some gas must leave the domain to carry away the angular momentum of the accreted material.  As $\dot{M}_\text{nozzle}$ is lowered, the fraction of material accreted onto the black hole increases.

The amount and duration of variability increase as $\dot{M}_\text{nozzle}$ is reduced.  This is in at least qualitative agreement with the early work of \cite{Spruit87}, who found that the effective $\al$-parameter induced by self-similar spiral shocks scales like $\mathcal{M}^{-3/2}$.  This implies that the viscous time for the disks, the typical time scale for large-scale evolution, is an increasing function of $\mathcal{M}$ and hence a decreasing function of $\dot{M}$.  Our setup is not self-similar, so we would not expect quantitative agreement.

\begin{figure}
\plotone{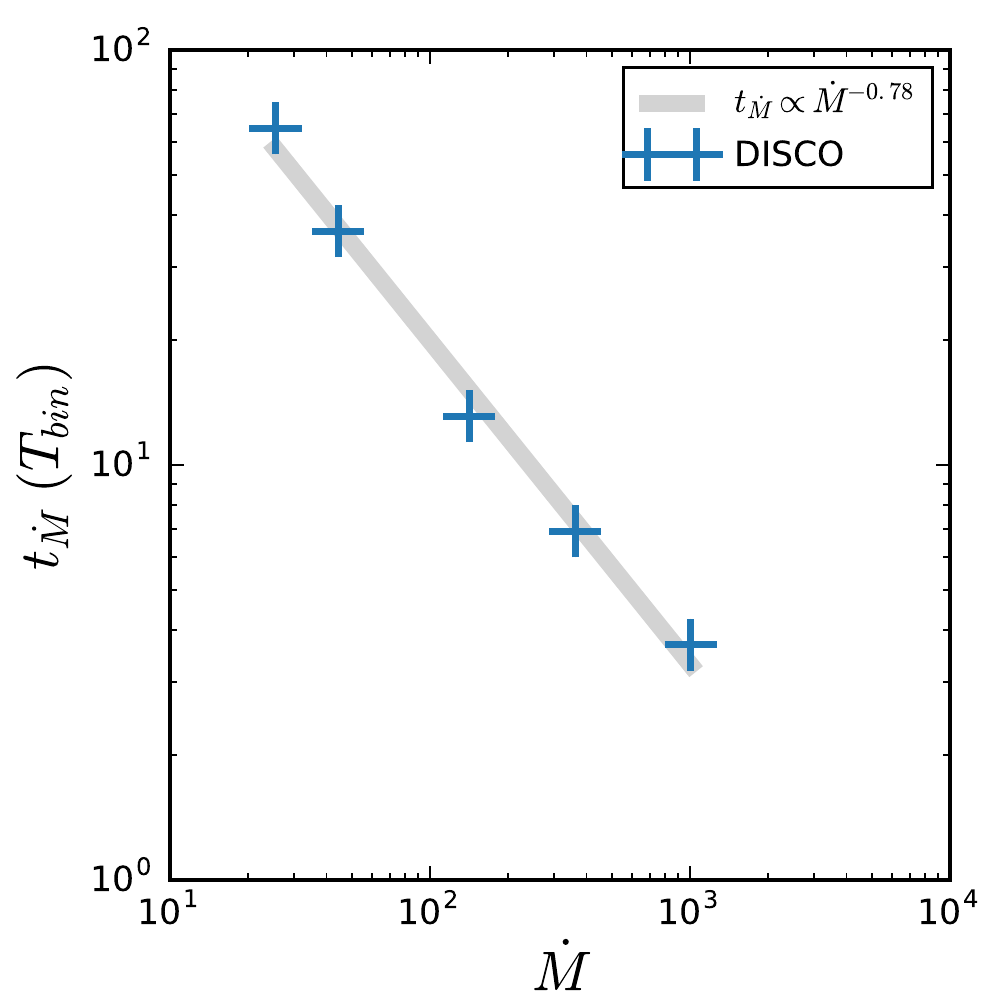}
\caption{\label{fi:tacc} Accretion time $t_{\dot{M}}=M_\text{disk}/\dot{M}$ as a function of $\dot{M}$ for all models (blue plus signs), and power-law fit to the simulation data (gray line).  Models \texttt{1}, \texttt{1.5}, and \texttt{2} can accrete their entire disk mass within the run time of the simulation ($29 T_\text{bin}$) and achieve quasi-equilibrium, unlike Models \texttt{2.5} and \texttt{3} with lower accretion rates. The dependence of $t_{\dot{M}}$ on $\dot{M}$ follows a rough power law of $-0.78 \pm 0.08$.}
\end{figure}

In Figure \ref{fi:tacc} we plot the accretion timescale $t_{\dot{M}} = M_\text{disk} / \dot{M}$ as a function of $\dot{M}$ for all primary models at the end of their runs.  The total disk mass $M_\text{disk}$ is calculated by integrating the lab-frame surface density $u^0 \Sig$ over the whole computational domain, and the accretion rates are taken from the inner boundary.  We see that Models \texttt{2.5} and \texttt{3} have accretion timescales longer than the runtime of the simulation, explaining the extended variability seen in Figure \ref{fi:mdot_all}.  A power-law fit to the data gives a slope of $-0.78 \pm 0.08$. This is of relevance to global Newtonian circumbinary accretion simulations, which typically employ a mass sink that removes some fraction of material every orbit.

\subsection{Wave Propagation}
\label{subsec:res-prop}

As a first diagnostic we check whether the two-armed spiral shocks adhere to the dispersion relation for tightly wound linear waves given by Equation \eqref{eq:dispRelPitch}.  The shock detector (see Section \ref{subsec:shockDet}) returns the azimuthal coordinates of the leading $\Phi_{i,\text{lead}}(r)$ and trailing $\Phi_{i,\text{trail}}(r)$ edge of each shock $i=A,B$.  We consider the center of each shock $\Phi_i = (\Phi_{i,\text{lead}} + \Phi_{i,\text{trail}}) / 2$.  The pitch angle of a spiral $\Phi(r)$ is
\begin{equation}
	\tan \theta = -\frac{1}{r \Phi'(r)} \ . \label{eq:deftanp} 
\end{equation}
We compute $\tan \theta_i$ for each shock according to Equation \eqref{eq:deftanp} by performing a numerical centered difference on $\Phi_i(r)$.  At each radius we compute the average Mach number $\mathcal{M}$ as
\begin{align}
	\langle \mathcal{M} \rangle &=  \frac{|u|}{c_s / \sqrt{1-c_s^2}} \ , \\
	\text{where: } |u| &= \sqrt{\gamma_{\mu\nu} \langle u^\mu \rangle  \langle u^\nu \rangle} \nonumber \ .
\end{align}

\begin{figure}
\plotone{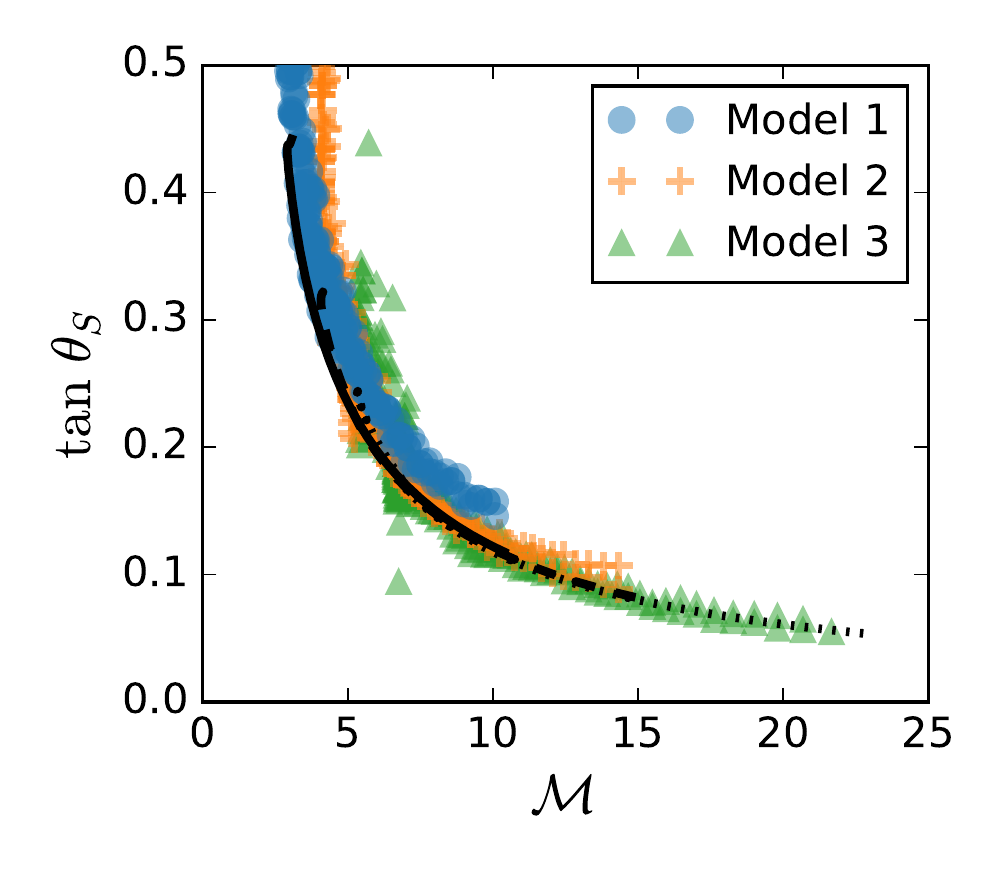}
\caption{\label{fi:disp} Pitch angle versus average Mach number for \model{1} (blue circles), \model{2} (orange plus signs), and \model{3} (green triangles). Lines are the analytic prediction from the dispersion relation for tightly wound linear waves for \model{1} (solid), \model{2} (dashed), and \model{3} (dotted).  Spiral shocks propagate in a nearly linear regime, with better agreement at higher Mach numbers and lower accretion rates.}
\end{figure}

In Figure \ref{fi:disp} we plot the pitch angle of shocks found in the minidisk simulations against the average Mach number of each annulus. We find good agreement between the numerically calculated pitch angles and the theoretical relationship $\eqref{eq:dispRelPitch}$, indicating that the shocks propagate mostly in the linear regime.

\subsection{Shock Dissipation}
\label{subsec:diss}

To quantitatively measure the coupling between the spiral shocks and the bulk disk flow, we measure the irreversible heating at each shock in post-processing.  Given the shock locations, we compute the jump in specific entropy $\Delta s_i = s_{i,\text{trail}} - s_{i,\text{lead}}$ over each shock at each radius.  To measure the specific irreversible heating over a shock of strength $\De s$ \cite{Rafikov16} defines the parameter $\psi_Q$ for a $\Gam$-law gas:
\begin{equation}
	\psi_Q \equiv \frac{1}{\Gam - 1} \left( e^{(\Gam -1)\De s } - 1 \right) \ . \label{eq:def-psi}
\end{equation}
Then the specific irreversible heating at a shock is simply $T_\text{lead} \psi_Q$.  To determine the irreversible heating rate at a shock, the specific heating rate must be multiplied by the mass flux through the shock surface $\Phi(r)$.  When summed over all the shocks in an annulus, one gets the total irreversible heating rate:
\begin{equation}
	\dot{Q}_{irr} = \sum_i \left[ \Sigma \left(r u^\phi - r \Phi'(r) u^r \right) T \right]_{i,\text{lead}} \psi_{Q,i} \ . \label{eq:QirrRaf}
\end{equation}

\begin{figure}
\plotone{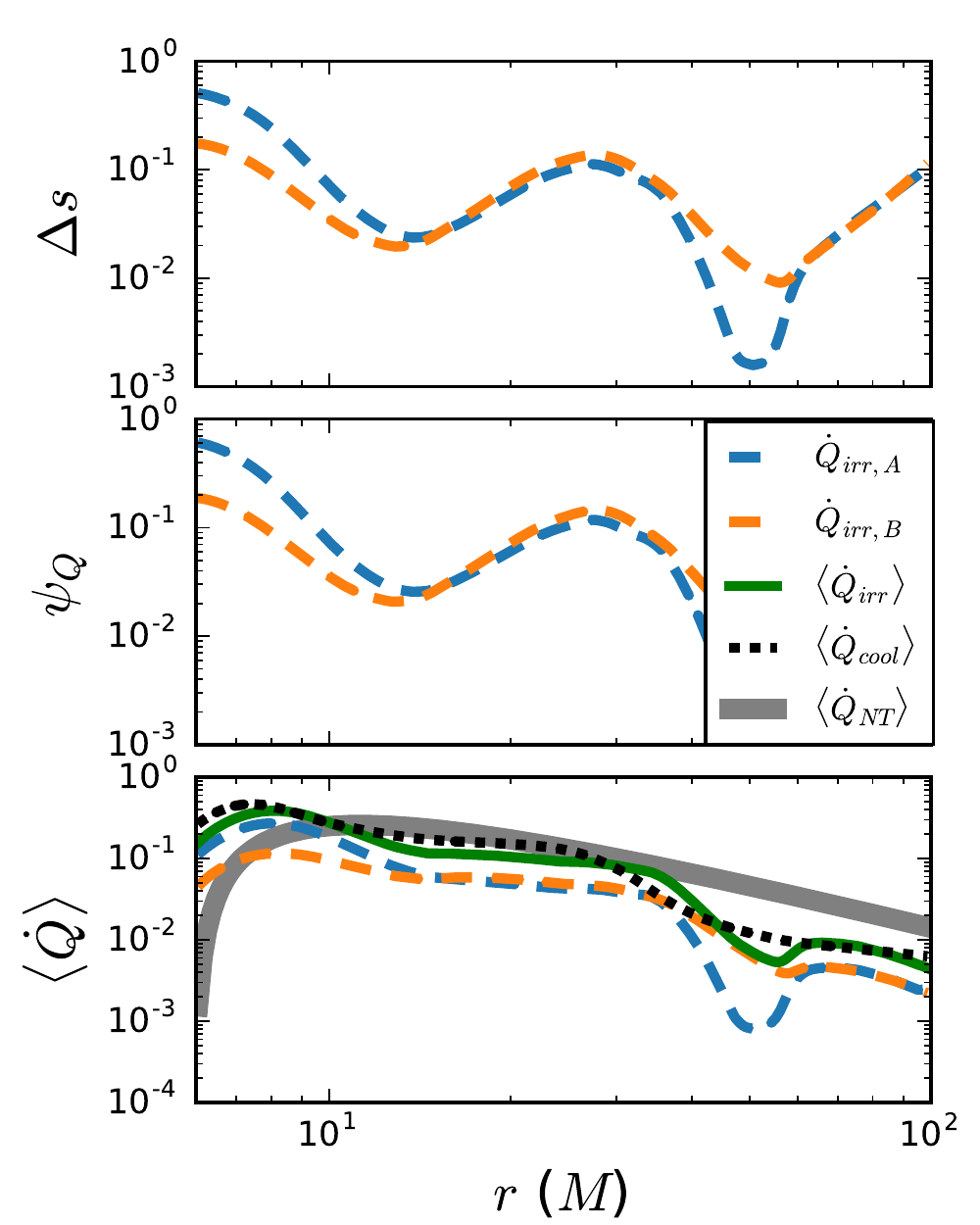}
\caption{\label{fi:diss} Radial profiles of dissipative shock quantities in \model{2}, time averaged from $(20 - 29) T_\text{bin}$.  Top to bottom: the entropy jump $\De s$, $\psi_Q$, and irreversible heating $\dot{Q}$. The shocks $A,B$ are plotted in blue and orange dashed lines, respectively.  The bottom panel also contains the total irreversible heating according to Equation \eqref{eq:QirrRaf} (green solid line), the local cooling rate integrated over each annulus $\avet{\dot{Q}_\text{cool}}$  (black dotted line), and the same quantity from a Novikov--Thorne disk with $\dot{M} = 0.75 \dot{M}_2$ in gray.  Shock-dissipated heat is nearly balanced by radiative cooling and roughly follows the Novikov--Thorne profile until the inner $r\lesssim 10M$.}
\end{figure}

Figure \ref{fi:diss} shows the radial profiles of $\Delta s_i$, $\psi_{Q,i}$, $\avet{\dot{Q}_\text{irr}}$, and $\avet{\dot{Q}_{cool}}$ for \model{2} averaged from $20 T_\text{bin}$ to $29 T_\text{bin}$.  Note that $\avet{\dot{Q}_\text{irr}}$ does not include heating (cooling) due to adiabatic compression (expansion) and advection.  Both $\De s$ and $\psi_Q$ vary over an order of magnitude through the disk, reflecting the double-peaked radial distribution seen in Figure \ref{fi:sig-fid-28}. On average, $\psi_Q \sim 0.1$.  

The distribution of irreversible heating and cooling in \model{2} has a much smoother radial dependence than $\psi_Q$. The is reflective of the universal character of emission from thin accretion disks: the profile of $\avet{\dot{Q}_\text{cool}}$ for a steady axisymmetric thin disk subject to local dissipation depends only on the accretion rate and black hole parameters, not the particular form of the dissipation or opacities.  For comparison we plot the expected Novikov-Thorne $\avet{\dot{Q}_\text{NT}}$ using the asymptotic value of $\dot{M}$ from Figure \ref{fi:mdot-fid}. 

The heating and cooling are broadly similar over the extent of the disk, with $\avet{\dot{Q}_\text{irr}} \lesssim \avet{\dot{Q}_\text{cool}}$ most often.  Even in a steady state one would not expect them to match exactly, as some energy must be advected inwards.  Both $\avet{\dot{Q}_\text{irr}}$ and $\avet{\dot{Q}_\text{cool}}$ agree with the Novikov-Thorne profile in $12 M \lesssim r \lesssim 30M$ but far exceed it in the inner disk as $r \to r_\text{ISCO}$, peaking at $r \approx 7.5M$.

\subsection{Angular Momentum Transport}
\label{subsec:angmom}

Dissipation in the fluid flow is a sign of angular momentum transport and hence accretion. Indeed in every simulation we find that the accretion rate through the inner boundary (into the black hole) matched the accretion rate through the outer boundary (matter injection from the nozzle minus outflow).  A quasi-equilibrium, where the inner and outer accretion rates match, typically occurred within a few ($<10$) $T_\text{bin}$. 

The relative strength of stress in an accretion flow is typically measured by determining an effective Shakura--Sunyaev (or Novikov--Thorne) $\al$-parameter.  The gravitational forces of the companion, plus the global character of the shocks providing the dissipation, make $\al_{\text{eff}}$ a somewhat imprecise notion.  We present two measures of $\al_{\text{eff}}$.  First, following \cite{Ju16}, we calculate the $\al_{\dot{M}}$ required at each radius for a Novikov--Thorne disk of the same average temperature and accretion rate.  Second, \cite{Rafikov16} determines for shocked Newtonian disks $\al_{\dot{Q}} \sim \psi_Q / 3\pi$.  We take this same prescription without modification, as $\psi_Q$ is a scalar quantity and hence frame independent.  

\begin{figure}
\plotone{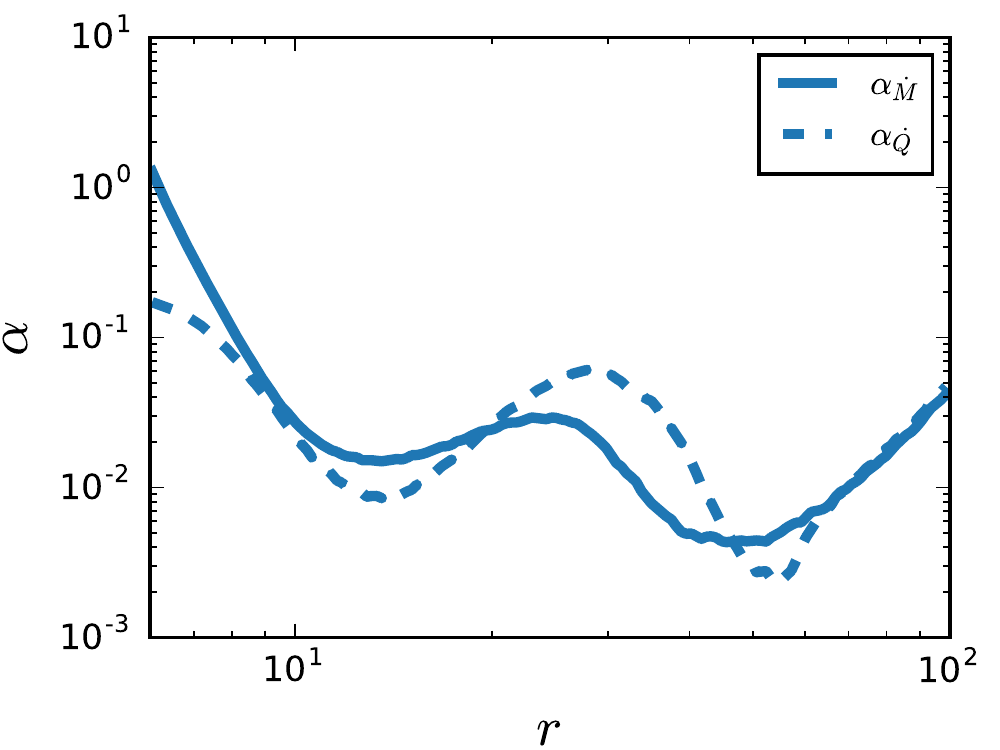}
\caption{\label{fi:alpha} Radial profile of effective $\al$ parameters in \model{2} measured from the accretion rate $\al_{\dot{M}}$ (solid) and the local dissipation $\al_{\dot{Q}}$ (dashed).}
\end{figure}

The radial profiles of both $\al_{\dot{M}}$ and $\al_{\dot{Q}}$ are plotted in Figure \ref{fi:alpha} for the fiducial run.  We see that they are broadly similar across the disk, except near the ISCO at $r=6M$ where $\al_{\dot{M}}$ diverges owing to the singularity in the Novikov--Thorne model.

To see the effect of the spiral shocks in detail, we decompose the torques acting on the disk according to Equation \eqref{eq:angMomDecom} (see Section \ref{subsec:angMomDecom} for details). Each of the torques $\tau_{\dot{M}}$, $\tau_\text{Re}$, $\tau_\text{ext}$, and $\tau_\text{cool}$ can be measured directly from \Disco{} output.  The Reynolds torque $\tau_\text{Re}$ in particular measures the angular momentum transport due to nonaxisymmetric features.  One can estimate the torque due to spiral shocks from $\psi_Q$ directly \citep{Rafikov16}:
\begin{equation}
	\tau_\text{shock} = \frac{\dot{Q}_\text{irr}}{v^\phi - \Om_P} \ . \label{eq:tauRaf}
\end{equation}
If angular momentum transport in minidisks is completely due to shocks, $\tau_{shock}$ should match $\tau_{Re}$.

Alternately, one can ask what sort of local angular momentum flux could account for the accretion and shock heating in an azimuthally averaged sense (such as an $\al$-viscosity).  Such a flux would be an addition $t^{\mu\nu}$ to the gas stress energy tensor.  Assuming, like the viscous stress tensor, $u_\mu t^{\mu\nu}_{non-id} = 0$ the energy and angular momentum conservation equations are modified to
\begin{align}
	&\partial_t \ave{\Sig \eps u^\mu} + \partial_r \ave{\Sig \eps u^r} = \ave{\Pi \nabla_\mu u^\mu} - \dot{Q}_\text{cool} + \ave{t^{\mu\nu} \nabla_\mu u_\nu} \ ,\\
	&\partial_t \ave{\Sig h u^0 u_\phi + t^0_\phi} + \partial_r \ave{\Sig h u^r u_\phi + t^r_\phi} = \ave{f_\phi} \ .
\end{align}
Identifying $\ave{t^{\mu\nu} \nabla_\mu u_\nu} \sim \ave{t^r_\phi \partial_r u^\phi}$ with $\ave{\dot{Q}_\text{irr}}$, one gets the relation
\begin{equation}
	\tau_{t} = \partial_r \ave{t^r_\phi} \approx \partial_r \left(\frac{\dot{Q}_\text{irr}}{\ave{\partial_r u^\phi}} \right)\ . \label{eq:tauLoc}
\end{equation}
Such a local model neglects the global structure of spiral shock waves but may be interesting on the basis of comparison,  especially since it is the root of the $\alpha$-prescription.

\begin{figure}
	\plotone{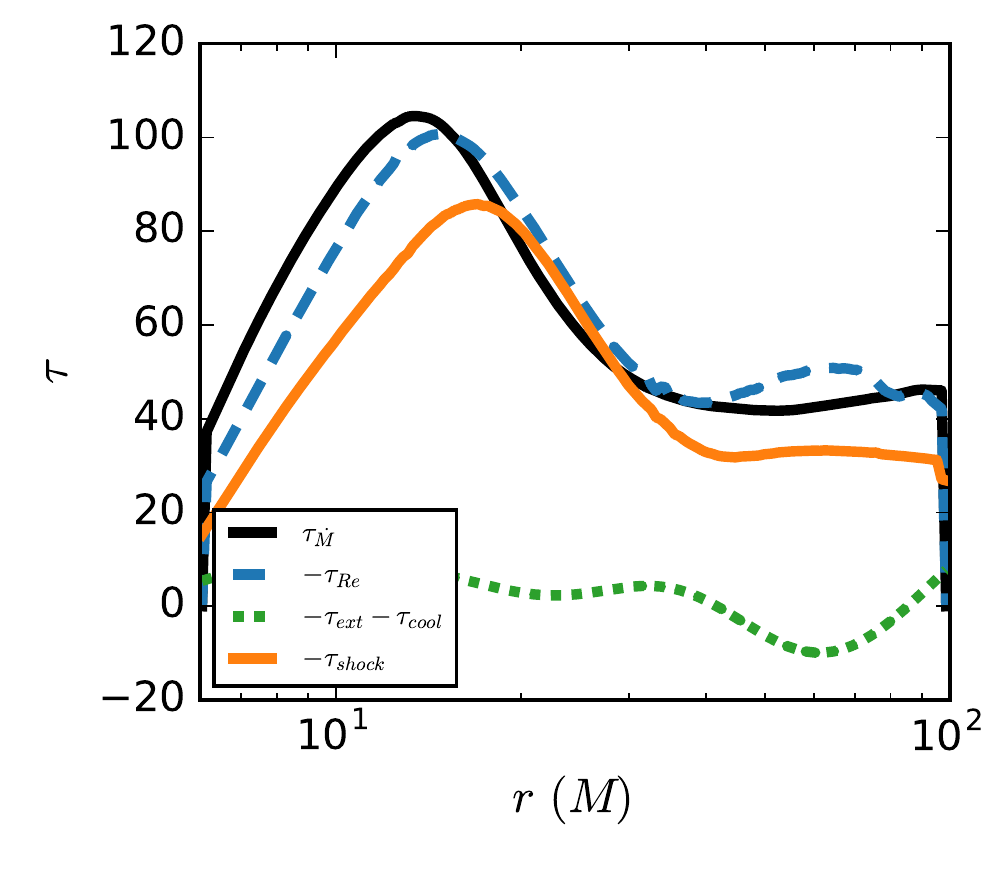}
	\caption{\label{fi:torque-m2} Time-averaged accretion torque $\tau_{\dot{M}}$ (black solid line), Reynolds torque $\tau_{Re}$ (blue dashed line), external and cooling torques $\tau_{ext}+\tau_{cool}$ (green dotted line), and predicted torque from spiral shocks $\tau_{irr}$ (Equation \eqref{eq:tauRaf}; orange solid line) for \model{1}. Time averaging was over $(20-29)T_\text{bin}$.  The predicted shock torques $\tau_\text{shock}$ follow the trend of the Reynolds torque $\tau_\text{Re}$.}  
\end{figure}

\begin{figure}
	\plotone{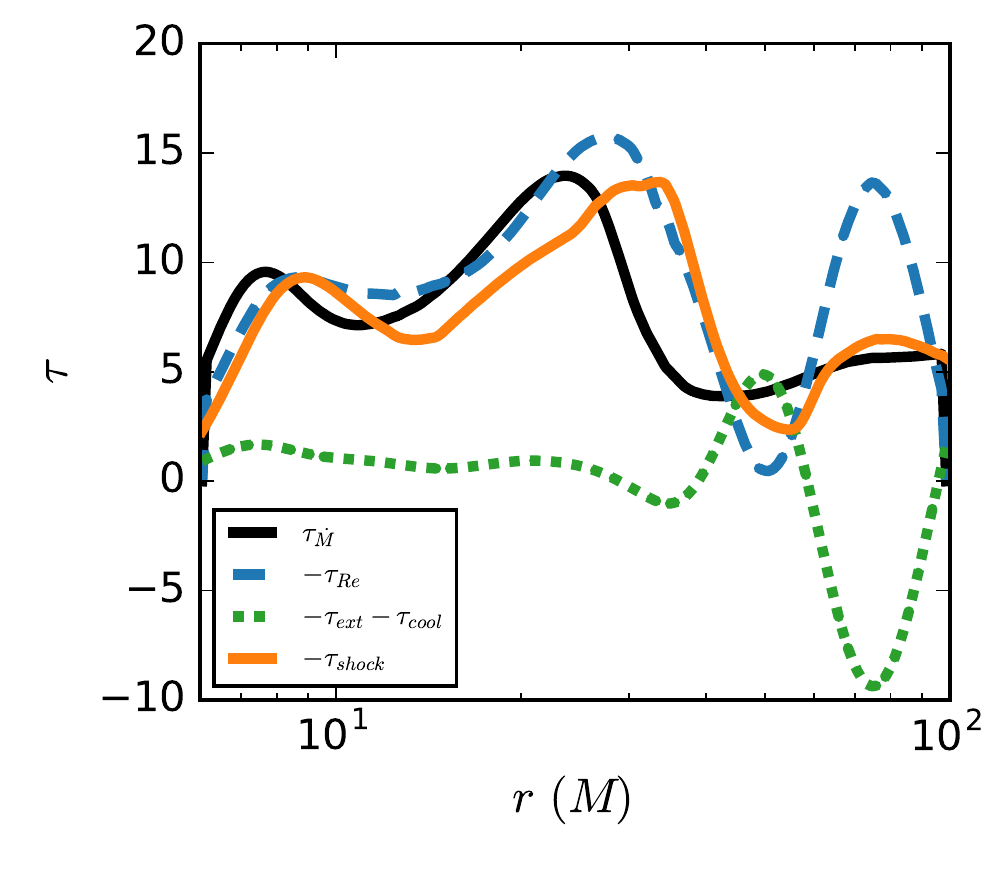}
	\caption{\label{fi:torque-m3} Same as Figure \ref{fi:torque-m2}, but for \model{2}}  
\end{figure}

\begin{figure}
	\plotone{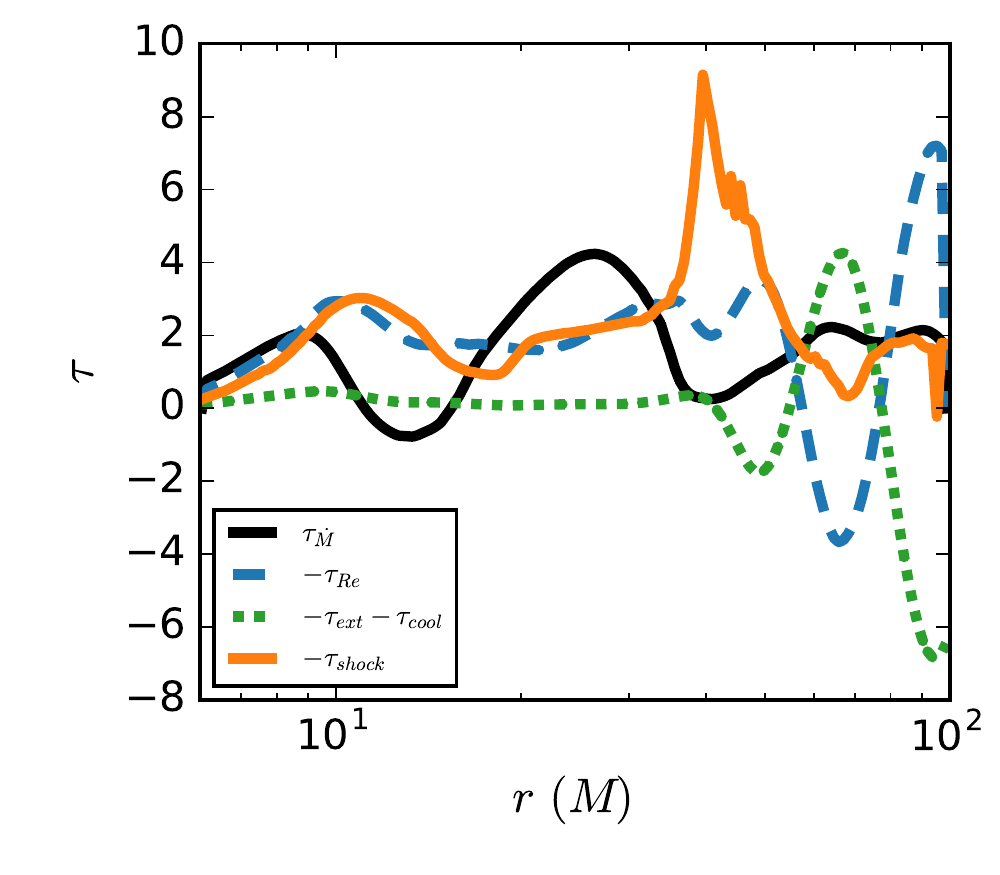}
	\caption{\label{fi:torque-m4} Same as Figure \ref{fi:torque-m2}, but for \model{3}}  
\end{figure}

We compare the predicted torque due to spiral shocks given in Equation \eqref{eq:tauRaf} to the measured $\tau_\text{Re}$ and $\tau_{\dot{M}}$ for the minidisk models in Figures \ref{fi:torque-m2}, \ref{fi:torque-m3}, and \ref{fi:torque-m4}.  The measured torques are calculated by azimuthally integrating grid quantities and centered differencing in $r$ when necessary.  The predicted torques due to irreversible shock heating, on the other hand, are calculated directly from the entropy jump $\De s$ measured to calculate $\psi_Q$.  

\model{1} is very near a steady state, with the accretion torque $\tau_{\dot{M}}$ balanced by $\tau_\text{Re}$, $\tau_\text{ext}$, and $\tau_\text{cool}$ throughout the $r < 100M$ region.  The small contribution of $\tau_\text{ext}$ and $\tau_\text{cool}$ to the torque balance indicates that the bulk of the angular momentum transport is provided by $\tau_\text{Re}$, a fact also true in Models \texttt{2} and \texttt{3}.  This alone indicates the non-axisymmetric structures drive accretion in these disks.  The discrepancy in the balance between $\tau_{\dot{M}}$ and $\tau_\text{Re}$ increases in \model{2} and again in \model{3}, indicating these disks are in much less steady states.  In \model{3} they widely differ, and there is even a narrow region undergoing deccretion.  

The shock torque model $\tau_\text{shock}$ shows good agreement with $\tau_\text{Re}+\tau_\text{ext}$, especially for Models \texttt{1} and \texttt{2} where it deviates by at most $30\%$ in $r < 60M$.  Even \model{3}, which is still evolving strongly in time, shows $\tau_\text{irr}$ agreeing well outside the band $40M \lesssim r \lesssim 50M$.  We find the local dissipation model $\tau_t$ does a poor job of predicting the torque on the disk.

\subsection{Spectral Comparison to Novikov--Thorne}
\label{subsec:spectra}

As seen in Figure \ref{fi:diss} the rate of cooling in the inner disk $r < 10M$ is in significant excess to the standard Novikov--Thorne model.  This naturally leads to the question as to whether there is an observational difference between a standard thin disk and one undergoing shock-driven accretion.  We calculate the observational spectrum of each model using the ray-tracing methods of Section \ref{subsec:raySpec}, based on the method of \cite{Kulkarni11}.

\begin{figure*}
	\plotone{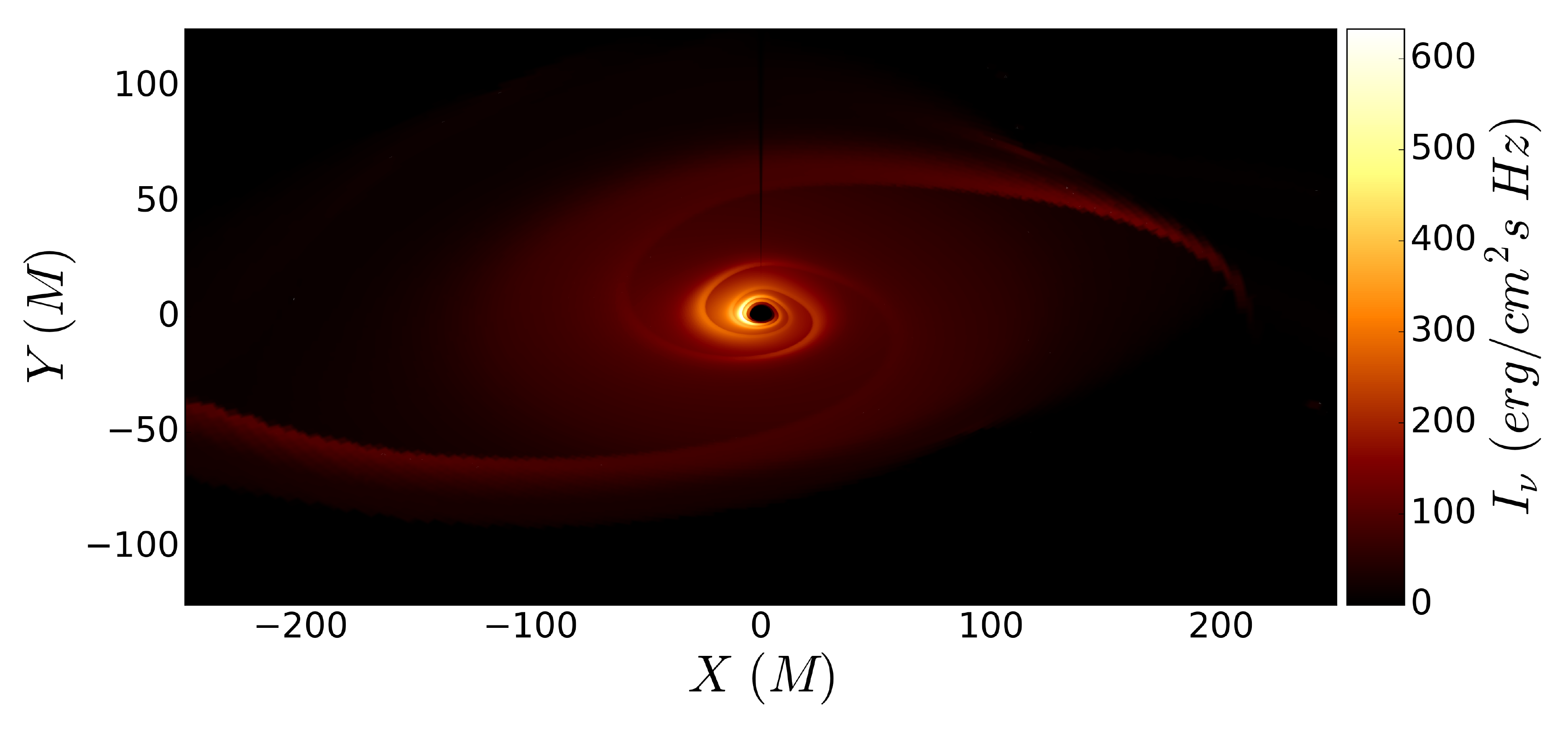}
	\caption{\label{fi:im} Ray-traced image of \model{2} at $\nu =   1$ keV. Shocks are clearly visible even near the black hole's shadow.  Relativistic beaming increases the intensity of observed emission in material with velocities toward the observer.}
\end{figure*}

\begin{figure*}
	\plotone{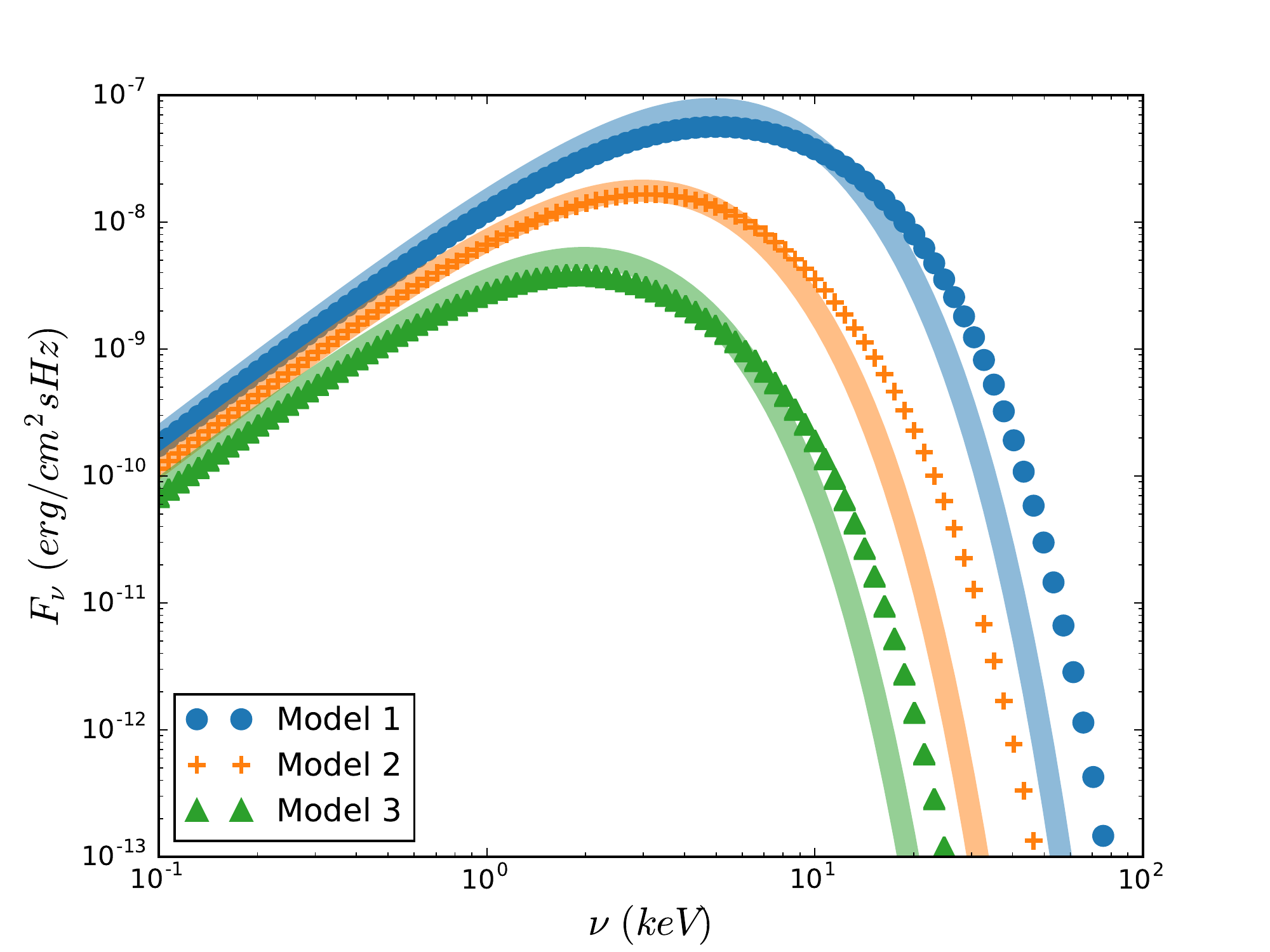}
	\caption{\label{fi:spec} Spectra of \model{1} (blue circles), \model{2} (orange plus signs), and \model{3} (green triangles) at $t = 28 T_\text{bin}$, obtained by integrating over the ray-traced intensity (e.g. in Figure \ref{fi:im}).  Solid lines are Novikov--Thorne spectra with $\dot{M}$ of the inner boundary averaged over $20 T_\text{bin}$ to $29 T_\text{bin}$ for each model.  Only the $R<40M$ region is included in the integration, to remove the effect of truncation at the outer disk edge.  The inclination angle $i=60^\circ$ and the distance $D=1$ kpc.  The data are consistent with the Novikov--Thorne profile at low energies but show a high-energy excess, consistent with the increased shock dissipation and radiative cooling observed at $r\lesssim 10M$.}
\end{figure*}

An image of the specific intensity $I_{0.1\text{ keV}}$ for \model{2} at $t=28 T_\text{bin}$ viewed at inclination $i=60^\circ$ is shown in Figure \ref{fi:im}.  The total emission is dominated by the bright inner region, and the two-armed structure is clearly visible. We generate the spectrum by integrating over such images at several frequencies.  The spectrum corresponding to Figure \ref{fi:im} is shown in Figure \ref{fi:spec}. 

The spectrum of each model at $28 T_\text{bin}$ is plotted in Figure \ref{fi:spec} with reference Novikov--Thorne curves.  Each spectrum is calculated including only the region $r<40M$, to ease comparison with the Novikov--Thorne model and between models by removing the effect of variable outer disk truncation.  Each Novikov--Thorne spectrum is calculated using the corresponding model's accretion rate at the inner boundary, time averaged over $(20-29)T_\text{bin}$.  In this comparison all model spectra are well characterized by the Novikov--Thorne spectrum, except for a slight excess at $\nu > 10$ keV.  This excess is precisely due to the large cooling rate at $r<10M$ seen in Figure \ref{fi:diss}, and corresponds to the presence of shock-heated gas near and within the ISCO.  

\subsection{Dependence on Inner Boundary Condition}
\label{subsec:bc}

Accretion onto a black hole has a well-defined physical inner boundary condition given by the event horizon.  This is advantageous for numerical studies, as it leaves no choices for the investigator. Simply extend the numerical grid through the horizon, and the metric will prevent any information about the inner edge of the grid from propagating outward into the observable simulation volume.  

Unfortunately, due to the CFL condition, this puts a heavy penalty on the time step for the hydro evolution.  Gas near the horizon plunges radially inward at almost the speed of light, so to capture the fastest modes in an explicit time evolution scheme, the time step must essentially be light limited. This makes long time evolution very computationally expensive and reduces the advantage of the azimuthal mesh motion in \Disco{}.

To evolve for several binary orbits, we find it necessary to move the inner boundary of the numerical grid to $r=4M$, above the event horizon.  The choice of $4M$ was made deliberately, as this is inside the sonic radius of the flow. Inside the sonic radius characteristics are all directed inward and the upwinded hydrodynamic evolution scheme prevents information from propagating outward.  This gives the same advantages as including the event horizon on the grid, but with a far smaller time step penalty.  The inner boundary condition within $4M$ is cold, isentropic inflow and matches the exterior solution very well.

\begin{figure}
	\plotone{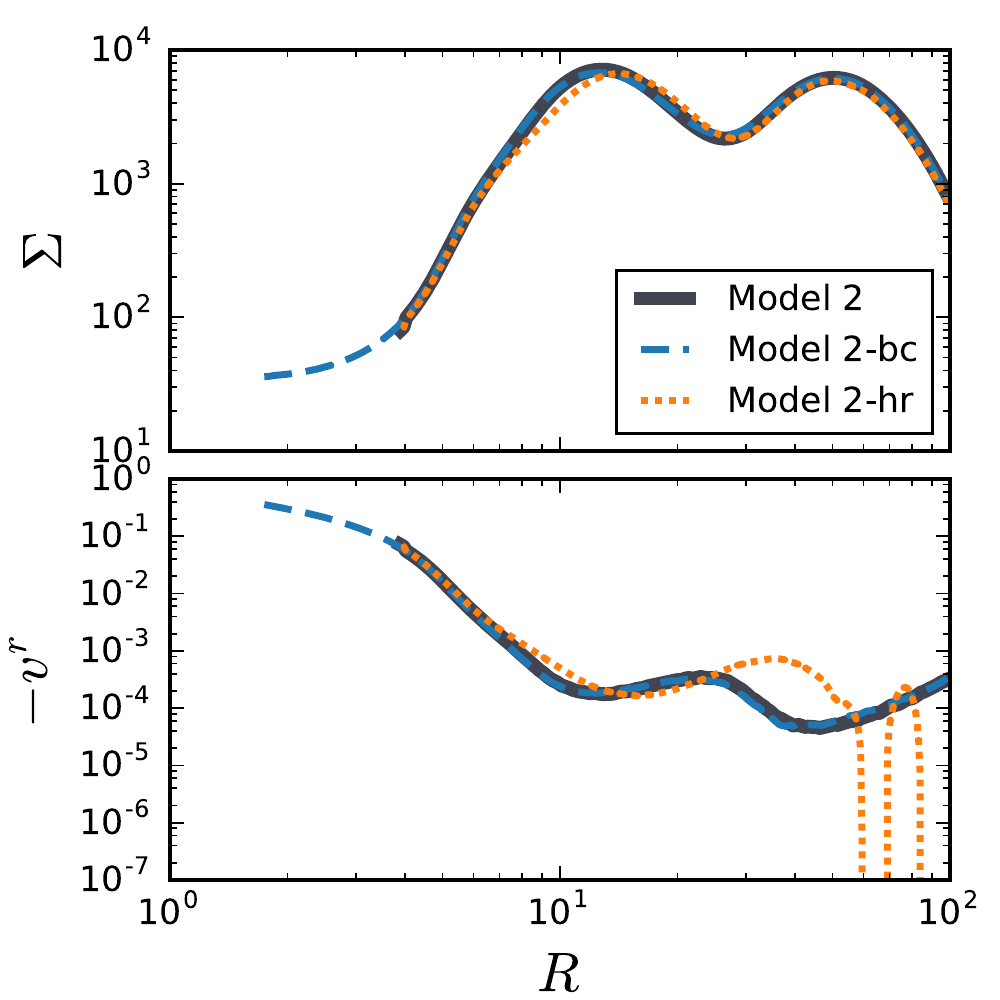}
	\caption{\label{fi:bc_hr_comp} Azimuthally averaged lab-frame surface density $\Sigma$ and radial velocity $\ave{v^r}$ for \model{2} at $29T_\text{bin}$ (solid gray line), \model{2-bc} at $26T_\text{bin}$ (dashed blue line), and \model{2-hr}  at $32T_\text{bin}$ (dotted orange line). \model{2-bc} smoothly continues \model{2} into the horizon even after an orbit of time evolution, indicating the robustness of the inner boundary condition.  The high-resolution \model{2-hr} develops time variability that does not settle down over the duration of the simulation, but the overall density distribution agrees well.}
\end{figure}

To verify that our boundary condition does not affect the global evolution, we mapped the \model{2} solution at $t=25 T_\text{bin}$ onto a new grid that does extend within the horizon.  This grid has $r_\text{min} = 1.8M$, and is referred to as \model{2-bc}.  We evolve \model{2-bc} for a single orbit, to $26T_\text{bin}$.  The azimuthally averaged surface density and radial velocity profiles of \model{2} and \model{2-bc} are shown in Figure \ref{fi:bc_hr_comp}.  The \model{2-bc} curves match those of \model{2} and smoothly continue through the horizon at $R=2M$.  This demonstrates the effectiveness of the isentropic inner boundary condition, even when placed above the horizon.

\subsection{Numerical Resolution}
\label{subsec:res}

To investigate the effects of numerical resolution, we mapped Model 2 at $t=28T_\text{bin}$ to a grid with double the resolution, 512 logarithmically spaced radial zones.  The simulation was run for four binary orbits, to $t = 32 T_\text{bin}$, and is denoted Model \texttt{2-hr}.

Model \texttt{2-hr} retains the strong two-armed shock structure. A similar clumpy accretion behavior develops as in the early evolution of \model{2} (between $8$ and $17$ $T_\text{bin}$, see Section \ref{subsec:fiducial}).  The remapping procedure upsamples the \model{2} grid, necessarily producing some small--scale noise.  This may be sufficient to push the model out of the quasi-equilibrium found by \model{2}. The clumpy accretion persists the length of the simulation, making a direct comparison between \model{2} and \model{2-hr} difficult.

The azimuthally averaged lab-frame surface density and radial velocity for \model{2-hr} are also plotted in Figure \ref{fi:bc_hr_comp}.  The data are from $t = 32T_\text{bin}$, where it appears that the clumpy accretion is beginning to settle down.  The unsteady nature of the flow is demonstrated in the radial velocity, where two regions can be seen with net positive velocity.  These fluctuations have little effect on the density distribution, which remains in good agreement with \model{2}.


\section{Discussion}
\label{sec:discussion}

We have studied the accretion dynamics onto a black hole in a binary system being fed by a circumbinary disk using 2D GRHD simulations. We have demonstrated the effectiveness of spiral shocks at driving accretion in these black hole ``minidisks,'' in broad agreement with recent Newtonian simulations in the context of CVs \citep{Ju16}  and circumplanetary disks \citep{Zhu16}.  The primary additions of this work are confirming the presence of shock-driven accretion when the disk is dynamically cooled, demonstrating that tidally driven spiral shocks can propagate throughout the disk and into the ISCO, enhancing emission of soft X-rays above the Novikov--Thorne model, and numerically verifying in the general relativistic case the relationship between disk torque and shock dissipation predicted by \cite{Rafikov16} for Newtonian disks.

We expect important aspects of our 2D simulations to carry over to the 3D case, though fully 3D GR(M)HD simulations of minidisks remain a topic for future study.  Analytic work has demonstrated that, while some wave modes are refracted out of the disk by the expected vertical temperature gradients, other modes will be channeled and amplified by the same gradients \citep{Lubow98}. \cite{Ju16} performed a single Newtonian 3D magnetohydrodynamic (MHD) simulation of a CV disk, where strong spiral shocks were clearly present and provided comparable torque to the MHD turbulence.  \cite{Bae16} performed a suite of Newtonian 3D disks, isothermal and adiabatic, subject to an $m=2$ perturbing potential.  They found all spiral waves were subject to an instability that disrupted the wave and generated turbulence.  The strength of the instability depended sensitively on the thermodynamics of the disk.

Clearly more work needs to be done to determine the role tidally induced spiral shocks play in accretion disks, both relativistic and nonrelativistic.  The results of \cite{Ju16}, \cite{Zhu16}, and \cite{Bae16} seem to indicate that spiral shocks can be dynamically important, either working in concert with the MRI and essentially providing a lower bound on $\al$, or by seeding turbulence directly.  The sensitive dependence on thermodynamics indicates that their ultimate role will be highly dependent on the cooling mechanisms and equations of state of the material at hand.

In black hole disks, if the shocks propagate undisturbed, they can provide dissipation at the ISCO, leading to a high-energy radiative excess over the standard Novikov--Thorne models.  This may be relevant for spin measurements of black hole X-ray binaries that rely on continuum fitting. Three-dimensional GRMHD simulations have also found MRI dissipation within the ISCO and conclude that continuum fitting based on Novikov--Thorne may mildly overestimate spins \citep{Penna10, Noble10, Kulkarni11, Schnittman15}.  A recent report raises the possibility of a shock in the outer, optically emitting region of LMC X-3 but found no similar evidence in the X-ray data \citep{Steiner14}.

Of relevance to SMBH binary detection, our results confirm that the minidisk SED resembles a standard disk blackbody apart from the high-energy excess (Figure \ref{fi:spec}).  Minidisks are soft X-ray sources undergoing orbital motion at potentially relativistic velocities.  In the Earth's frame, relativistic beaming will induce a periodic variation in each minidisk's emission that should be absent in the emission from the global circumbinary disk \citep{DOrazio15}.  Discriminating the minidisk emission from the circumbinary disk may be possible if the SED of the entire system has a ``notch,'' as suggested in \cite{Roedig14}.  However, global Newtonian simulations with \Disco{} show that gas thrown off the minidisks can impact and shock-heat the cavity wall, creating hot spots that can smooth the spectrum \citep{Farris15A}.

Emission of GWs shrinks the orbit of the binary, leading to merger in a time given by Equation \eqref{eq:Tmerge}, thousands of orbits for the parameters in this work.  We find that spiral shock patterns establish themselves very rapidly, in less than a single orbit of the binary, and can lead to a quasi-steady state in dozens of binary orbits.  This separation of timescales implies that spiral shocks will remain in quasi-steady state and drive accretion as the binary orbit secularly evolves.  This should remain the case until the binary begins evolving on an orbital timescale, when the separation is tens of $M_\text{bin}$. In this regime our quasi-Newtonian treatment of the binary potential ceases to be valid and a fully general relativistic binary black hole metric should be employed as in \cite{Farris12,Noble12,Zilhao15}.

After the LIGO discovery of GW150914 and GW151226, electromagnetic counterparts to stellar mass black hole binaries are of great interest \citep{LIGO16GW150914Discovery, LIGO16GW151226}.  If a black hole binary exists in a gaseous environment, for instance, within an AGN disk \citep{Bartos16, Stone17} or post-supernova fallback \citep{Perna16}, it will undergo circumbinary accretion and minidisks will form, potentially producing distinct radiative signatures.


\section{Summary}
\label{sec:summary}

We performed general relativistic hydrodynamical simulations of accretion disks in the Schwarzschild metric subject to tidal forces of a binary companion.  These disks, referred to as ``minidisks,'' were fed from a nozzle at the L2 Lagrange point, modeling the accretion streams feeding minidisks seen in circumbinary accretion simulations.  The tidal forces excited two-armed spiral shock waves that propagate throughout the disk,  generating heat through dissipation, transporting angular momentum, and efficiently driving accretion into the black hole.

The spiral shocks propagate primarily in the nearly linear regime, agreeing with the relativistic generalization of the WKB dispersion relation for tightly wound linear waves.  Measurements of the jump in specific entropy across the spiral shocks provide a measure of the irreversible heating of the disk.  The cooling profile qualitatively follows the Novikov--Thorne profile in the outer disk but maintains a significant excess of emission within $r \lesssim 10 M$.  The angular momentum transport in all models is driven by the Reynolds stress due to the spiral shocks, in agreement with the shock-driven torque model of \cite{Rafikov16} to within $\sim30\%$.  The stress corresponds to an effective Shakura--Sunyaev $\al$-parameter on the order of a few$\times 10^{-2}$.

Accretion via spiral shocks is a purely hydrodynamical effect, occurring without the need for magnetic fields or radiation.  Any disk in a binary system will be subject to tidal forces, and the ensuing spiral shocks can carry a non-negligible portion of the angular momentum transport budget.  Spiral shocks can propagate through the ISCO and may provide an emission excess over the standard Novikov--Thorne models.


\section{Acknowledgements}
We are grateful to Patrick Cooper, Dan D'Orazio, Paul Duffell, Brian Farris, Andrei Gruzinov, Zoltan Haiman, and Roman Rafikov for many useful conversations.  This work was supported in part by NASA through Astrophysics Theory (ATP) Grant NNX11AE05G.  Computations were performed on the Pleiades cluster at NASA AMES and on the Ria cluster at the Center for Cosmology and Particle Physics, New York University.


\bibliography{minidisc_sources}



\end{document}